\documentclass[twocolumn,showpacs,preprintnumbers,amsmath,amssymb,pre,,superscriptaddress]{revtex4-1}
\usepackage[utf8]{inputenc}

\usepackage{xcolor}    
\usepackage{graphicx}
\usepackage{dcolumn}
\usepackage{bm}
\usepackage{subfigure}
\usepackage{amssymb}
\usepackage{multirow}
\usepackage{mathtools}
\mathtoolsset{showonlyrefs=true}

\usepackage{braket}
\usepackage{placeins}

\graphicspath{{plots/}}

\usepackage{braket}
\usepackage{upgreek}

\usepackage{tikz}
\usepackage{upgreek}
\usetikzlibrary{shapes,backgrounds,calc,patterns}
\tikzset{subfiglabel/.style={left, minimum size=4ex}}%

\renewcommand{\pi}{\uppi}

\begin{document}


\title{Ionization potential depression and Fermi barrier in warm dense matter--a first--principles approach}

\author{Michael Bonitz}
\author{Linda Kordts}
\affiliation{
 Institut f\"ur Theoretische Physik und Astrophysik, Christian-Albrechts-Universit\"at zu Kiel,
 Leibnizstra{\ss}e 15, 24098 Kiel, Germany}

\begin{abstract}
Ionization potential depression (IPD) is of crucial importance to understand and accurately predict the properties of dense partially ionized plasmas. Many models of IPD have been developed that, however, exhibit largely varying results.  Recently, a novel approach was proposed that is based on first-principles quantum Monte Carlo simulations and that was demonstrated for hydrogen [Bonitz \textit{et al.},  Phys. Plasmas (2024)]. Here this concept is discussed in more detail. Particular attention is devoted to the Fermi barrier that electrons have to overcome upon ionization and which significantly contributes to IPD when the nuclear charge increases. 
\end{abstract}

\pacs{xxx}

\maketitle
\section{Introduction}\label{s:intro}
In a partially ionized plasma the number of free electrons, $N_e$, is related to the number of ions via the Saha equation,
\begin{align}
    \frac{N_e N_{a+1}}{N_{a}} &= 2V \left( \frac{2\pi mk_BT}{h^2} \right)^{3/2}\frac{\sigma_{a+1}}{\sigma_a} \; e^{- I^{\rm EK}_a/k_BT}\,,
    \label{eq:saha-ek}\\
    I^{\rm EK}_a & \equiv \epsilon_{a+1}(1)-\epsilon_{a}(1)\,,
     \label{eq:deltai-ek}
\end{align}
where, to set the basis, we have reproduced an expression given by G.~Ecker and W.~Kröll in Ref.~\cite{ecker-kroell_63}. In Eq.~\eqref{eq:saha-ek}, $N_a$ denotes the number of $a$-fold charged ions and $\sigma_a$ its partition function of the internal motion. Further, $ I^{\rm EK}_a $, in Eq.~\eqref{eq:deltai-ek} denotes the ionization energy of the $a$-ion which is defined as the difference of the ground state energies of ``ion a+1'' and ``ion a''. In the case of the hydrogen ionization balance, $a \to 0$ and $N_a \to N_A$, is the number of atoms, whereas $N_{a+1} \to N_p$ is the number of protons. Note that this quantity depends on the definition of the zero energy level. Taking the lowest value of the proton energy as a reference, $\epsilon_p = 0$, it follows $I^{\rm EK} = -\epsilon_{1s} = + 1$Ryd $\ge 0$, and the number of free electrons following from Eq.~\eqref{eq:saha-ek} is exponentially suppressed with $I^{\rm EK}$. 

It is well known that, in a dense plasma, individual atoms are affected by the surrounding medium -- through Coulomb interaction, as well as quantum and exchange effects -- which results in a lowering of the binding energy below the isolated atom limit. According to Ecker and Kröll \cite{ecker-kroell_63}, Eq.~\eqref{eq:saha-ek} retains its form and only the exponent gets modified, 
\begin{align}
I^{\rm EK} \to I^{\rm eff}(n_e;T) = I^{\rm EK} + \Delta I(n_e;T)    \,,
\label{eq:ieff-ek}
\end{align}
where the ionization energy is replaced by an effective one that acquires a shift, $\Delta I$ -- the ionization potential depression (IPD) --  that depends on electron density and temperature, cf.~Eq.~\eqref{eq:ieff-ek}. $\Delta I$ is not only required to understand the behavior of atomic bound states in a plasma, it is also used to infer important plasma properties, including density, temperature, equation of state or opacity.

The ionization potential depression, $\Delta I$, has been studied theoretically for a long time. In addition to Ref.~\cite{ecker-kroell_63}, early works include those of R.~Rompe and M.~Steenbeck \cite{rompe-steenbeck_39}, D.R.~Inglis and E.~Teller \cite{inglis-teller_39}, J.C.~Stewart, and K.D.~Pyatt,~Jr.~\cite{stewart-pyatt_66}, and B.J.B.~Crowley~\cite{crowley_pra_90}. 
These are phenomenological models that either neglect part of the relevant effects, such as quantum and spin effects, continuum lowering, strong Coulomb correlations, shift of bound state energies, and excited atomic states,   or treat them in a simplified manner. Due to their simplicity, the models of Stewart and Pyatt (SP) and Ecker and Kröll (EK) are very popular and particularly broadly applied to systems with multiple ionic charge states such as aluminum or carbon, e.g., \cite{preston_hedp_13, ciricosta_prl_12}.
Also, recently an improved chemical model has been presented \cite{yuan_pre_25}.
In addition to models for dense plasmas, extensive investigations have been performed for the lowering of the binding energy of excitons in optically excited semiconductors.  in the framework of the Bethe-Salpeter equation of non-equilibrium Green functions theory \cite{zimmermann_pssb_78,haug_pqe_84, red-book}. The difficulty here is to self-consistently treat quantum and dynamical screening effects in the bound state problem, for a recent overview, see Ref.~\cite{roepke_pre_19}.

Recently, experimental results for IPD or continuum lowering, for different materials have been reported. Vinko \textit{et al.} reported creation of solid-density aluminum using an x-ray FEL \cite{vinko_nat_12}. They detected K-$\alpha$ fluorescence following K-shell photoabsorption for a set of x-ray photon energies, that was affected by IPD.
Ciricosta \textit{et al.} extended this experiment to measurements of the position of the K edge of dense aluminum for six photon energies and observed significantly better agreement with the EK model compared to the SP model~\cite{ciricosta_prl_12}, for an analysis of the models, see Ref. \cite{kasim_np_18}. Additional measurements of the K edge position for various elements and compounds that were isochorically heated by x-rays were reported in Ref.~\cite{ciricosta_ncom_16}.
Doeppner \textit{et al.} investigated beryllium that was compressed at the NIF to pressures up to 3 Gbar and temperatures around $2\cdot 10^6$K \cite{doeppner_nature_23} and observed pressure ionization of the K edge. From XRTS spectra they deduced the average charge state and degree of ionization.
Finally, we mention recent transient absorption measurements of dense copper using ultrashort XFEL pulses by Mercadier \textit{et al.}~\cite{mercadier_np_24}.

On the other hand, recently theoretical investigations appeared, that are based e.g., on average atom models \cite{militzer_prr_22, santra_pre_22, huang_pre_24,davletov_njp_23} and density functional theory \cite{hu_prl_17, militzer_prr_22, santra_pre_22, huang_pre_24, vinko_nat-com_14}.
Furthermore, S.X.~Hu predicted another interesting effect for dense carbon at low temperatures that he called ``Fermi surface rising'', i.e., a possible blue shift of the ionization energy at high density \cite{hu_prl_17}.

The difficulty of many models is that IPD results from the interplay of many factors and requires to accurately account for the interaction among charges (electrons, ions), neutrals (atoms, molecules) and between both. This makes a consistent treatment of IPD a difficult theoretical problem. Furthermore, all atomic bound states are affected by the plasma simultaneously, which requires a systematic thermodynamic approach.

Here we consider a different approach that is based on first principles fermionic path integral Monte Carlo (PIMC) simulations for hydrogen \cite{filinov_pre_23} and was briefly introduced in Ref.~\cite{bonitz_pop_24}. We discuss this approach in more detail, starting from a general derivation of the Saha equation, fully including quantum effects and excited atomic bound states. In order to understand the capabilities and limitations of this method, we concentrate on the simplest case of hydrogen and then discuss extensions to $Z$-fold charged hydrogen-like ions.

This paper is organized as follows: In Sec.~\ref{s:theory} we re-derive the mass action law for a hydrogen plasma with medium effects included. In Sec.~\ref{s:degeneracy} we extend the theory to quantum degenerate electrons. In Sec.~\ref{s:ipd-hydrogen} we present our first-principle numerical results for the IPD. The case of Z-fold charged ions is discussed in Sec.~\ref{s:ipd-zfold}.

 \section{Mass action law}\label{s:theory}

 \subsection{Chemical potentials}
The mass action law is derived from the thermodynamic stability condition of chemically reacting spatially uniform systems and is expressed in terms of the chemical potentials of all species,
\begin{align}
    \mu_a = \mu^{\rm id}_a + \Delta\mu_a\,,
    \label{eq:chem-pot-def}
\end{align}
where $\mu^{\rm id}_a$ is the ideal chemical potential whereas $\Delta\mu_a$ denotes the interaction contribution. For fermions, the ideal chemical potential is given implicitly by the normalization,
\begin{align}
    n_a &= \frac{g_a}{\Lambda_a^3} I_{1/2}(x_a)\,,\quad g_a = 2s_a+1\,,\label{eq:fermi-normalization}\\
    I_\nu(x) &= \frac{1}{\Gamma(\nu+1)}\int_0^\infty dt \frac{t^\nu}{e^{t-x}+1}\,,\quad x_a=\beta\mu_a^{\rm id}\,,
\end{align}
where $g_a$ is the statistical weight (spin degeneracy) of species ``a'', and $I_\nu$ is the Fermi integral of order $\nu$. The spin degeneracy factors for electrons, protons (spin $1/2$ particles) and atoms (for the case of hydrogen, composite bosons consisting of an electron and a proton) are
\begin{align}
    g_e &=g_p=2\frac{1}{2} + 1 = 2\,,\label{eq:ge}\\
    g_A &= g_e g_p = 4\,.\label{eq:ga}
\end{align}
Note that the degeneracy factors can be quite complicated for larger atoms. However, for parameters where quantum effects of the ions can be neglected, we don't expect that the values of $g_i$ and $g_A$ have a measurable effect, as long as the relation \eqref{eq:ga} is satisfied.

Solving Eq.~\eqref{eq:fermi-normalization} for the ideal chemical potential, we obtain
\begin{align}
    \beta\mu_a^{\rm id} &= I_{1/2}^{-1}(\tilde\chi_a)\,,\quad \tilde\chi_a = \frac{\chi_a}{g_a}\,
    .\label{eq:ideal-chem-pot}\\
    \chi_a &= n_a\Lambda_a^3\,,\\
    \Lambda_a &= \frac{h}{(2\pi m_a k_BT)^{1/2}}\,,
\end{align}
where $\chi_a$ is the degeneracy parameter, $n_a=N_a/V$ the particle density, and $\Lambda_a$ the thermal de Broglie wavelength. 

In the non-degenerate limit the Fermi integral has a power series expansion in terms of the fugacity, $z=e^x$, with the leading order $I_\nu(z)=z$, resulting in $n_a=\frac{g_a}{\Lambda_a^3} e^x$, with the limit of \eqref{eq:ideal-chem-pot} giving the solution for the ideal non-degenerate chemical potential
\begin{align}
    \beta\mu_a^{\rm id}|_{\rm class} &= \ln \tilde\chi_a\,,
    \label{eq:muid-classical}
\end{align}
 
The mass action law was derived by Saha et al. more than 100 years ago \cite{saha_20} and applied to the computation of stellar spectra. However, the resulting traditional Saha equation neglects quantum effects of the electrons and internal atomic states and effects of the surrounding medium; it, therefore, has to be generalized, e.g. \cite{schlanges-etal.95cpp}.

We apply the chemical picture, following the presentation of Ref.~\cite{blue-book} where, in addition to free electrons and protons, bound pairs of electrons and protons are treated as independent species -- atoms in the ground state and excited states, with internal quantum number $j=n_j l_j m_j$. The corresponding bound state energies in the presence of the plasma are the solutions of the Bethe-Salpeter equation \cite{blue-book},
\begin{align}
    E_j(n_e,T) &= E_n^0 + \Delta_j(n_e,T)\,,\\
               &= \frac{E_{n=1}}{n^2} + \Delta_{nl}(n_e,T)\,,     
    \label{eq:ej-renormalized}
\end{align}
where $E_1 \equiv E_{n=1} =- 1$Ryd, and $\Delta_{nl}$ are the plasma-induced shifts of the atomic energy levels (selfenergy shifts) that depend on density and temperature.
Treating the atoms as new (nondegenerate) species  gives rise to chemical potentials that, in addition to  Eqs.~\eqref{eq:chem-pot-def} and \eqref{eq:muid-classical}, contain a contribution of the relative energy of the bound pair,
\begin{align}
     \beta\mu_{Aj} &= \ln \tilde\chi_{Aj} + \beta E_j + \beta\Delta \mu_{Aj}\,.
     \label{eq:mu-atom-def}
\end{align}
Note that the effect of all interactions on the bound states is included in the renormalized energy eigenvalues, cf. Eq.~\eqref{eq:ej-renormalized}, therefore, $\Delta  \mu_{Aj}$ contains only the renormalization of the center of mass part of the chemical potential. This part is missing in Ref.~\cite{blue-book}, and we will also assume that it is negligible and put $\Delta  \mu_{Aj} \to 0$, in the following.

 \subsection{Saha equation for a nonideal classical \\hydrogen plasma}

Recall the ionization equilibrium in a spatially uniform partially ionized plasma, which is derived from minimizing the free energies with respect to the species concentrations,
\begin{align}
    \mu_e + \mu_p = \mu_{A_{j1}}=  \mu_{A_{j2}}= \dots\,,
    \label{eq:mal}
\end{align}
We start by evaluating the ionization equilibrium between free electrons and protons (assuming  that all particles are nondegenerate) and atoms in an arbitrary bound state ``j'' \cite{blue-book}, by inserting the chemical potentials into the mass action law, Eq.~\eqref{eq:mal}. The result is 
\begin{align}
    \frac{n_e}{2}\Lambda_e^3\cdot \frac{n_p}{2}\Lambda_p^3\cdot e^{\beta(\Delta\mu_e+\Delta\mu_p)} = \frac{n_{Aj}}{4}\Lambda_A^3\cdot e^{\beta E_j}\,.
\label{eq:saha-statej}
\end{align}
Reordering and simplifying this expression and, taking into account that $\Lambda_p^3 \approx \Lambda_A^3$, we obtain
\begin{align}
    \frac{n_{Aj}}{n_en_p}\, e^{-\beta(\Delta\mu_e+\Delta\mu_p)} = \Lambda_e^3 \,e^{-\beta E_j}\,.
\end{align}

\subsubsection{Saha equation for the ionization equilibrium}
Since we are not interested in the population of the individual bound states (this would be relevant, e.g. for optical properties), we now sum over all states $j$ and introduce the total atomic number density, 
$n_A = \sum_j n_{Aj}$. In addition, we introduce a new notation for the density of free electrons, $n_e^*$, which gives the total density of electrons according to $n_e=n_e^* + n_A$. Finally, we take charge neutrality into account, leading to $n_p^*=n_e^*$. The summation yields, for the classical case, 
\begin{align}
    \frac{n_A}{n_e^{*2}}\bigg|_{\rm class} &= \Lambda_e^3\, \Sigma_A^{\rm int}\:e^{\beta \Delta I_0 } \equiv K_A\,,\label{eq:saha}\\
    \Sigma_A^{\rm int} &= \sum_j e^{-\beta E_j} = \sum_{nl}(2l+1) \, e^{-\beta (E_n^0+\Delta_{nl})}\,,\label{eq:sigma-a-int}\\
    \Delta I_0 &= \Delta\mu_e+\Delta\mu_p\,,\label{eq:cont-shift}\\
    K_A(n_e,T) &= \Lambda_e^3\,\sum_{nl}(2l+1) \, e^{\beta I^{\rm eff}_{nl}(n_e,T)}\,,\label{eq:ka-def}\\
    I^{\rm eff}_{nl}(n_e,T) &= |E^0_{n}| + \Delta I_{nl}(n_e,T)\,,\label{eq:ieff-nl}\\
    \Delta I_{nl}(n_e,T) &= \Delta I_0(n_e,T) - \Delta_{nl}(n_e,T)\,.
    \quad\label{eq:delta-i-nl}
\end{align}
Equation~\eqref{eq:saha} is the familiar Saha equation for ionization and recombination in a nonideal hydrogen plasma with nondegenerate particles and generalizes expression \eqref{eq:saha-ek}. In Eq.~\eqref{eq:saha}, the r.h.s. contains the internal partition function $\Sigma_A^{\rm int}$, including all bound states in the plasma medium, and an exponential factor that is independent of the quantum numbers of the bound states. The exponential contains an energy shift, $\Delta I_0$, that is determined by the interaction parts of the chemical potentials of the free particles and contributes to the continuum lowering, as will be discussed in Sec.~\ref{sss:ipd}. Further, in the r.h.s. of Eq.~\eqref{eq:saha} we introduced the mass action constant, $K_A$, which allows us to rewrite the internal partition function in terms of effective ionization energies, Eq.~\eqref{eq:ieff-nl},  of all bound states that generalize the ground-state expressions of Eq.~\eqref{eq:ieff-ek}. However, they still neglect quantum degeneracy effects that will be included in Sec.~\ref{sss:ipd}.

\subsubsection{Regularization of the partition function}\label{sss:planck-larkin}
It is well known that, for a hydrogen atom (or similar systems with Coulomb bound states), the canonical partition function over the bound states, $Z_A^{\rm int} = \sum_{n=1}^\infty n^2 e^{-\beta E_n}$, diverges. If continuum states are taken into account properly, this problem disappears. One way to achieve a convergent result that, moreover, is continuous when the density is changed is to use the Planck-Larkin partition function, where the two lowest expansion terms of the exponent are being subtracted, which follows from Levinson's theorems, e.g. \cite{blue-book}
\begin{align}
    Z_A^{\rm PL} &= \sum_{nl} (2l+1) \left\{e^{-\beta E_{nl}} - 1 + \beta E_{nl}\right\}\,,
    \label{eq:pl-function}\\
    K_A^{\rm PL} &= \Lambda_e^3\sum_{nl} (2l+1) \left\{e^{\beta I^{\rm eff}_{nl}} - 1 - \beta I^{\rm eff}_{nl}\right\}\,,\label{eq:pl-ka}
\end{align}
where we used the renormalized energy eigenvalues \eqref{eq:ej-renormalized} and introduced the superscript ``PL'' to denote the Planck-Larkin procedure. In the following, we will use $Z_A^{\rm PL}$, instead of $Z_A^{\rm int} $, and $K_A^{\rm PL} $, instead of $K_A$. Note that in the case of atoms in a plasma medium, the number of bound states is always finite; therefore, also the original partition function could be used, as it is convergent. However, a disadvantage of this choice is that it leads to discontinuities when the effective ionization potential changes.

\subsubsection{Dimensionless form of the Saha equation}\label{s:dimless}
We now introduce the degree of ionization  and the (dimensionless) atom fraction,  
\begin{align}
\alpha(n_e,T) &=\frac{n_e^*(n_e,T)}{n_e}\,,\\
x_A(n_e,T) &= \frac{n_A(n_e,T)}{n_e}\,.
\end{align}
If the plasma contains only free electrons, protons, and atoms, then $\alpha+x_A=1$. If, in addition, molecules are present, our approach is applicable as well, but then  $\alpha$ and $x_A$ have to be regarded as independent quantities. The same holds if other bound states are present. 
In our first-principles approach both $\alpha$ and $x_A$ are known input, so it is sufficient to consider the ionization equilibrium in order to solve for the IPD.

We now rewrite the classical expression, Eq.~\eqref{eq:saha}, by introducing the species fractions, 
\begin{align}
    \frac{x_A}{\alpha^2}\bigg|_{\rm class} &= \chi_e\, 
    \Sigma_A^{\rm PL}e^{\beta (\Delta\mu_e+\Delta\mu_p)} = n_e K^{\rm PL}_A(n_e,T)\,.
\label{eq:saha-dimless}
\end{align}
This equation still neglects quantum degeneracy effects of the electrons which we now restore.

\section{Saha equation for degenerate electrons. Definition of IPD, continuum lowering, and Fermi barrier}\label{s:degeneracy}
\subsection{Quantum and spin statistics effects in the Saha equation}
So far we have neglected the Fermi statistics of the electrons. However, these effects are important, for the case $\chi_e\gtrsim 1$, and this requires us to use, for electrons, the ideal chemical potential, Eq.~\eqref{eq:ideal-chem-pot} where we need to replace $\chi_e \to \chi_e^*$, which depends on the free-electron density, $n_e^*$, that is, $\chi^*_e=\alpha\chi$. Then, we obtain from Eq.~(\ref{eq:saha}), the more general Saha equation
\begin{align}
    \frac{n_A}{n_i^*} &= \frac{x_A}{\alpha}= 2 \, \Sigma_A^{\rm PL} \, e^{\beta \mu_e^{\rm id,F}(\alpha \cdot \tilde\chi_e)} e^{\beta (\Delta \mu_e+\Delta \mu_p)}\,. \label{eq:saha-fermi}
\end{align}
This equation goes over to the classical Saha equation \eqref{eq:saha-dimless}, at low densities, when $\chi_e \to 0$. Equation~\eqref{eq:saha-fermi} is the proper starting point to compute the degree of ionization if the interaction contributions of the chemical potentials, $\Delta \mu$, the continuum lowering, and the renormalized bound-state energies, $E_{nl}$, are known.

\subsection{Definition of the ionization potential depression, continuum lowering, and Fermi barrier}\label{sss:ipd}
Equation \eqref{eq:saha-fermi} contains all medium influences that arise from Coulomb interaction, quantum and spin effects, however, it does not provide direct access to the effective binding energies and the associated ionization potential depression. In order to find the proper generalizations of Eqs.~\eqref{eq:ieff-nl} and \eqref{eq:delta-i-nl}, for the quantum case,
 it is necessary to bring Eq.~\eqref{eq:saha-fermi} to the form of Eq.~\eqref{eq:saha-dimless}, as it corresponds to the classical low-density limit. This is achieved by dividing by $\alpha$, with the result
\begin{align}
    \frac{x_A}{\alpha^2} &= 
    \chi_e \Sigma_A^ {\rm PL}\,e^{\beta \Delta I^{\rm F}_0(\tilde\chi^*_e)}
    \label{eq:pl-function-fermi}
    \equiv n_e K^{\rm PL,F}_A\,, 
    \\
\Delta I^{\rm F}_0(n_e,T) & \equiv  \Delta I_0(n_e,T) + \Delta I^{\rm F}(\tilde\chi^*_e) \,,\label{eq:def-i0f}\\
 K^{\rm PL,F}_A & \equiv e^{\beta\Delta I^{\rm F}(\tilde\chi^*_e)}
    \, K^{\rm PL}_A\,,\label{eq:kpla-fermi}\\
 \Delta I^{\rm F}(\tilde \chi_e) &= \mu_e^{\rm id,F}( \tilde\chi_e)-k_BT \ln \tilde\chi_e\,,\label{eq:delta_i-fermi}   
\end{align}
and we introduced the superscript ``F'' to denote the Fermi statistics. 
Compared to the low density limit, Eq.~\eqref{eq:saha-dimless}, the partition function in Eq.~\eqref{eq:pl-function-fermi} has remained unchanged, but in the exponential prefactor, the energy shift $\Delta I_0$, Eq.~\eqref{eq:cont-shift}, is replaced by $\Delta I^{\rm F}_0$, Eq.~\eqref{eq:def-i0f}. Similarly, the mass action constant is replaced by a more general result for a quantum (fermion) system~\footnote{Note that in order to include this contribution into all energy levels, we first have to go back to the original partition function and make the transition to the Planck-Larkin form at the final step.}, $K^{\rm PL,F}_A$, Eq.~\eqref{eq:kpla-fermi}.
The exponential factors in Eqs.~\eqref{eq:pl-function-fermi} and \eqref{eq:kpla-fermi} contain a novel quantity, $\Delta I^{\rm F}$, which will be called ``Fermi barrier'', and is given by the difference between the fermionic and classical ideal chemical potentials of the electrons. The physical meaning of this quantity will be discussed in more detail in Sec.~\ref{sss:ipd-disucssion}.

Finally, the fermionic mass action constant \eqref{eq:kpla-fermi} allows us to identify the renormalized ionization potentials. Instead of Eq. \eqref{eq:ieff-nl}, that holds for low density, they are given by the more general expression for fermions
\begin{align}
    I^{\rm eff,F}_{nl}(n_e,T) &=  |E_n^0| + \Delta I^{\rm F}_{nl}(n_e,T)\,,
    \label{eq:ieff-nl-fermi}   \\
     \Delta I^{\rm F}_{nl}(n_e,T) & = \Delta I^{\rm F}_0(n_e,T) - \Delta_{nl}(n_e,T)     \,.
\label{eq:ipd-fermi}
\end{align}
Here, $I^{\rm eff,F}_{nl}$ is our final result for the effective ionization energy of the orbital with quantum numbers, $n,l,m$, whereas $\Delta I^{\rm F}_{nl}$, Eq.~\eqref{eq:ipd-fermi}, denotes the shift of the ionization energy of this level, compared to the single-atom case, i.e. the ionization potential depression (IPD). This shift contains an orbital-dependent contribution $\Delta_{nl}$, denoting the shift of the bound state level (relative to the low-density limit) and a second contribution that is independent of the quantum numbers. As we will show in Sec.~\ref{sss:ipd-disucssion}, this term $\Delta I^{\rm F}_0(n_e,T)$, Eq.~\eqref{eq:def-i0f} is identified as the position of the continuum of free electron-proton pairs and is often termed ``continuum lowering'', even though this may be misleading. 
 
\subsection{Discussion of the shift of the continuum, of IPD  and over the Fermi barrier ionization}\label{sss:ipd-disucssion}
\subsubsection{IPD due to Coulomb interaction}
\begin{figure}
    \centering    
    \includegraphics[angle=0,width=0.5\textwidth]{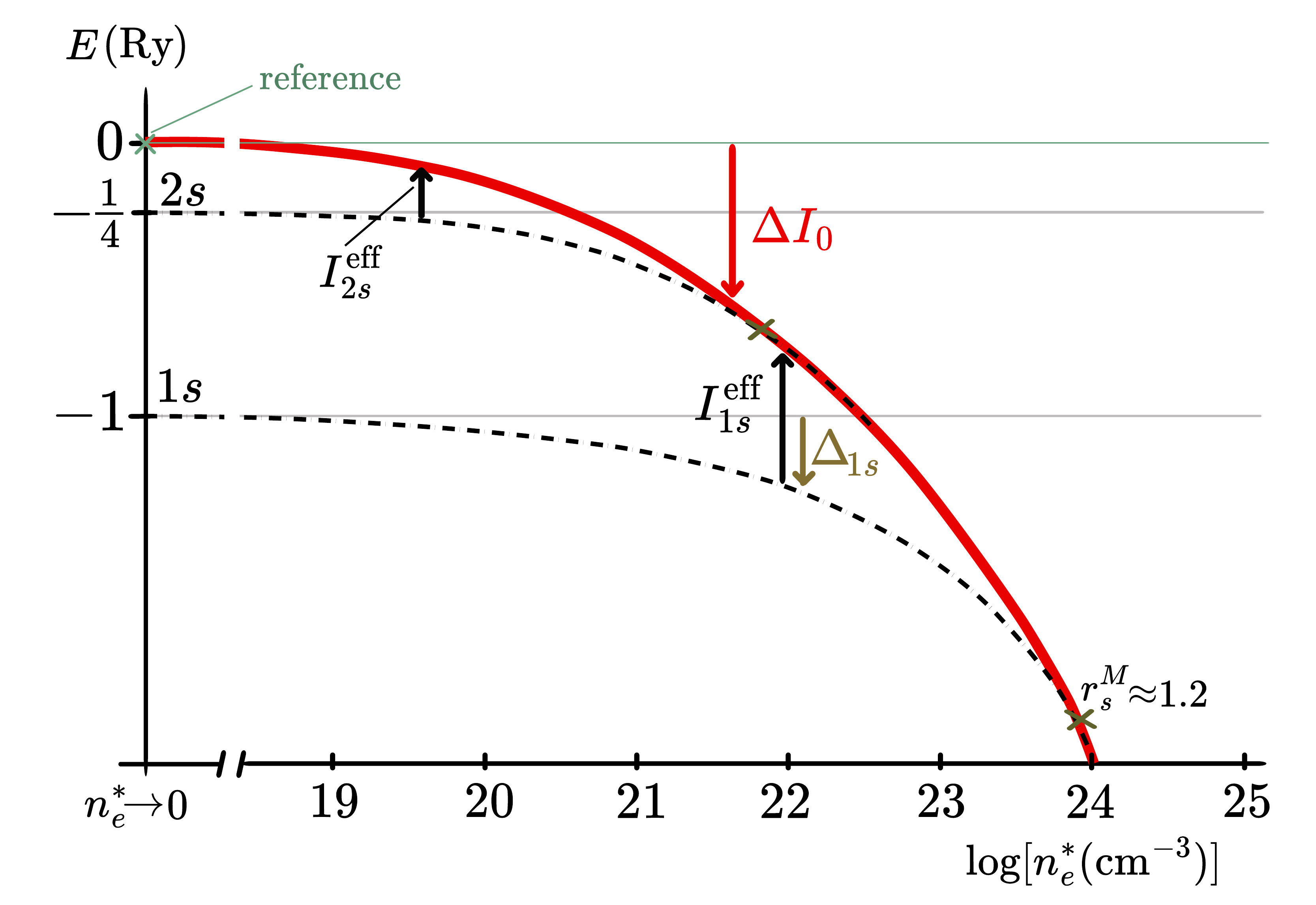}
    \caption{Illustration of the standard picture of the bound state spectrum in a plasma of free electron density $n^*_e$, following from solving the Bethe-Salpeter equation. Dashes: (negative) shifts of the 1s- and 2s-levels, denoted $\Delta_{1s}$ and $\Delta_{2s}$. Red curve: shift of the zero energy level, $\Delta I_0$ (``continuum''), without including $\Delta I^{\rm F}$. The effective ionization energies are sketched by the vertical black arrows. The energy reference point (zero value) is the zero density position of the continuum. }
    \label{fig:demo1}
\end{figure}

While there exist many different definitions of the IPD and continuum position, in our approach there is no ambiguity. The obvious energy reference point for both, $\Delta I^{\rm F}_0$ and $\Delta_{nl}$, is the low-density (i.e., isolated atom) limit, corresponding to $n_e\to 0$, where both quantities vanish exactly. This is illustrated in Fig.~\ref{fig:demo1}.
The effective ionization energies, $I_{nl}^{\rm eff}$, are given by the distance from the renormalized binding energy, $E_{nl}=|E_n^0|-\Delta_{nl}$, to the renormalized continuum, $\Delta I_0^{\rm F}$, cf. the black vertical upward arrows in Fig.~\ref{fig:demo1}, for the examples of the 1s- and the 2s-states. 

Note that, for the ionization energies, $I_{nl}^{\rm eff}$, and the IPD, which can be directly measured in a photoabsorption experiment, the reference point does not play a role (it is the same for all bound-state levels and for the continuum). On the other hand, photoabsorption does not give direct access to 
the position of the bound states and of the continuum. For an understanding of the physics, however, a separate computation of both quantities is important. We, therefore, rewrite the position of the continuum, Eq.~\eqref{eq:def-i0f}, explicitly,
\begin{align}
    \Delta I^{\rm F}_0(n_e,T) &= 
\Delta I_0(n_e,T)
    + \Delta I^{\rm F}(\tilde\chi^*_e)\,.\quad\label{eq:def-i0f-2}
\end{align}
Let us start with a discussion of the first term on the r.h.s. It describes the contributions to the chemical potentials that arise from interactions of an electron (and proton) with all other particles of all ``chemical species'', including electrons, protons, and atoms. 

\begin{figure}
    \centering
    \includegraphics[angle=-0,width=0.48\textwidth]{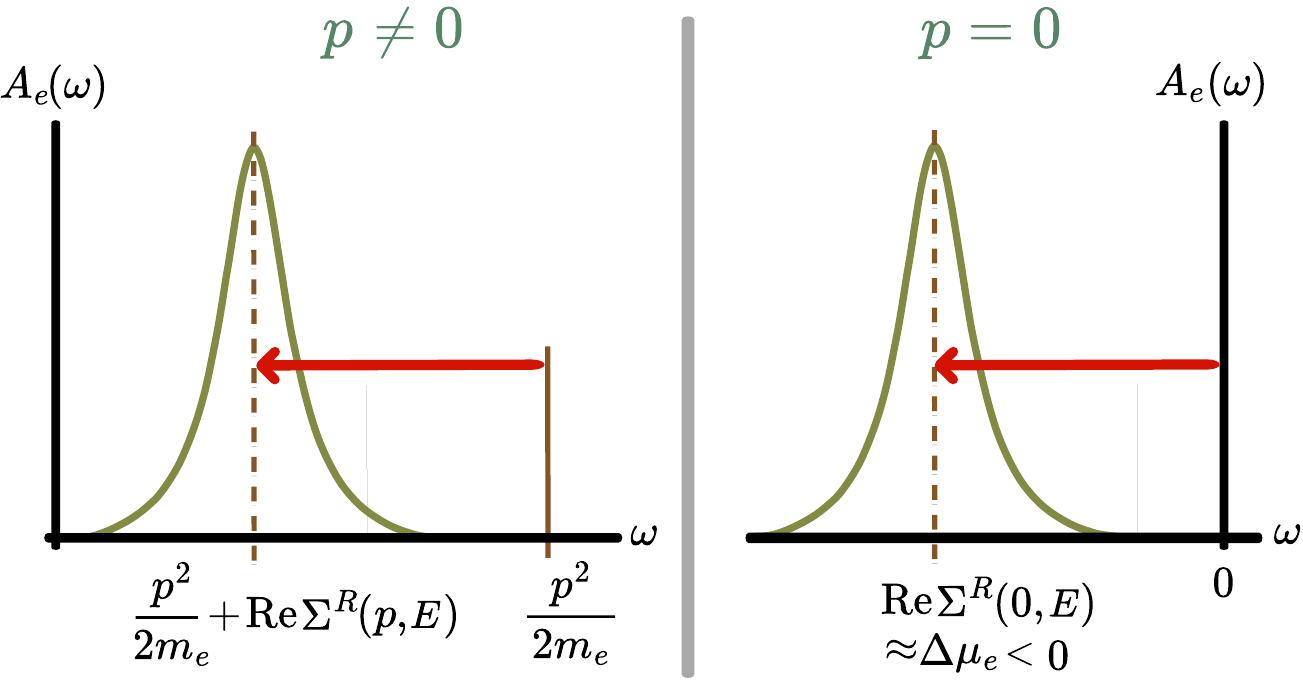}
    \caption{Illustration of the spectral function of unbound electrons. In the quasiparticle approximation, for an electron with momentum $p$, the spectral function is peaked at the single-particle energy, $E(p)=p^2/2m + {\rm Re}\Sigma(p,E)$. The selfenergy shift (red arrows) can be approximated, for $p=0$ (continuum edge), by the interaction part of the chemical potential (``rigid shift approximation'' \cite{blue-book}) which takes into account interaction contributions between the electron and all other particles. The peak width is given by Im$\Sigma$ and corresponds to a finite quasiparticle lifetime resulting from correlations. }
    \label{fig:demo2}
\end{figure}

On the other hand, in a physical picture, where no distinction is being made between free and bound electrons \cite{bonitz_pop_24}, the chemical potentials have to be understood as the energy required to create an electron (and a proton). This addition energy depends on the momentum of the particles and is, therefore, directly related to the density of states and the spectral function, $A(p,\omega)$. An illustration is shown in Fig.~\ref{fig:demo2}, for finite and zero momentum, respectively. For a given momentum, $p$, the quasiparticle energy, in the presence of the plasma, is $E(p)= p^2/2m+$Re$\Sigma(p,E)$, where $\Sigma$ denotes the selfenergy of Green functions theory, e.g.,~\cite{blue-book,bonitz_qkt}.
In this approach, the chemical potential follows from the spectral function by momentum averaging, where the kinetic energy gives rise to the ideal chemical potential, while the interaction part, $\Delta\mu_e$, is related to the real part of the self-energy. From this, the position of the continuum of electron-proton pairs is easily identified: for free particles, it corresponds to the momentum, $p=0$ and $E^{\rm free}(0)=0$. On the other hand, in the plasma, the continuum position shifts to $E(0)= $Re$\Sigma(0,E)$, which is negative; cf. the red line in Fig.~\ref{fig:demo1}.

\subsubsection{Over the Fermi barrier ionization (OFBI)}
Note that the continuum shift, $\Delta I_0$, is due to correlations. However, it neglects exchange effects which are present in a quantum system, even if the particles of the medium would be non-interacting. To ionize an atom, in a dense plasma, it is not sufficient to overcome the ionization potential: in addition, the electron needs to find an empty quantum state in the continuum. Due to the Pauli principle of fermionic particles, low-lying states may be highly occupied, giving rise to an additional energy barrier -- the Fermi barrier. We, therefore, call this effect ``Over the Fermi barrier ionization (OFBI)'', as it closely resembles a similar process in strong-field laser ionization. In fact, for photoionization of atoms or molecules to be successful, the photon should not only provide the ionization potential, but the ionized electron must, in addition, gain kinetic energy (the ponderomotive potential) sufficient to catch up with the oscillating laser field \cite{kremp_99_pre,bonitz_99_cpp}. This process has been called ``Above threshold ionization (ATI)'' or ``over the barrier ionization'' and was first discussed by Agostini \textit{et al}.~\cite{agostini_prl_79}, for recent reviews, see Refs.~\cite{becker_aamop_02,agostini_jpb_24}. 
Therefore, the complete expression for the continuum position, Eq.~\eqref{eq:def-i0f-2}, contains, in addition to the interaction contribution, the Fermi barrier, $\Delta I^{\rm F}$, Eq.~\eqref{eq:delta_i-fermi}. 

\begin{figure}
    \centering
    \includegraphics[angle=0,width=0.48\textwidth]{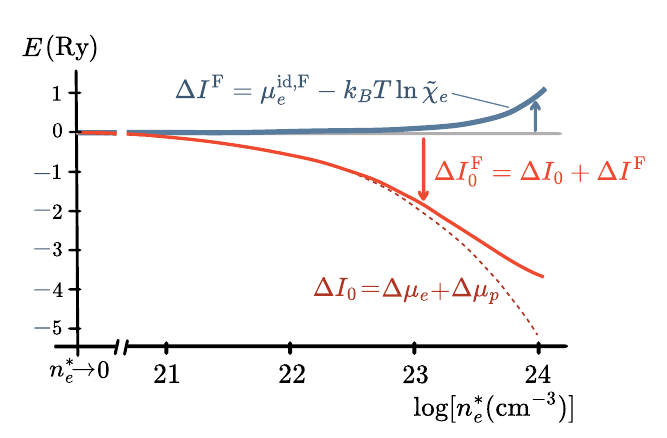}
    \caption{Illustration of the total shift of the continuum edge, $\Delta I_0^{\rm F}$ (the lowest energy state of an unbound electron-proton pair). It equals zero at zero density, which is used as energy reference. At finite density, this energy level (``continuum'') experiences a down-shift, due to interactions (red dashed line, corresponding to $\Delta I_0$ shown in Fig.~\ref{fig:demo1}), but also an up-shift,  by $\Delta I^{\rm F}$, due to the Fermi statistics among all fermions -- the Fermi barrier. At $T=0$ and high density, the latter is approximately given by the Fermi energy, $E_F$.}
    \label{fig:demo3}
\end{figure}

While the correlation-induced selfenergy shift of the continuum edge is negative, the Fermi barrier, $\Delta I^{\rm F}$, is positive. The importance of $\Delta I^{\rm F}$ is particularly clear at low temperatures and high densities where the electrons behave as a nearly ideal Fermi gas. Then, an additional electron in the continuum cannot be created with zero momentum but, due to the Pauli principle, the minimum energy (momentum) state that is available corresponds to $E=E_F$ ($p=p_F$). With increasing density this energy increases as $n_e^{2/3}$, giving rise to an up-shift of the continuum edge that competes with the correlation-induced down-shift shown in Fig.~\ref{fig:demo1}. The consequences for the total continuum shift are sketched in Fig.~\ref{fig:demo3}.

A Fermi statistics-induced up-shift of the continuum was introduced for dense low-temperature carbon by S.X. Hu \cite{hu_prl_17} and was termed ``Fermi surface rising''. In that paper, a density-induced up-shift of the carbon 1s-level, $\epsilon_{1s}$ was reported, and the K-edge location was computed, by adding to $|\epsilon_{1s}|$, the Fermi energy, $E_{\rm K edge} = |\epsilon_{1s}| + E_F$. Translating this into our notation, the first term can be identified with a part of the effective ionization energy, $I_{1s}^{\rm eff}$, as defined in Eq.~\eqref{eq:ieff-nl},
\begin{align}
   |\epsilon_{1s}| & = |E^0_{1s}|-\Delta_{1s}+\Delta I_0 \,,
\end{align}
but the correction due to the Fermi statistics (Fermi barrier), in general, should be $\Delta I^{\rm F}$, instead of $E_F$, cf. Eq.~\eqref{eq:ieff-nl-fermi}. Only at zero temperature and extremely high densities the result reduces to the Fermi energy. Thus, our result is a generalization of Ref.~\cite{hu_prl_17} to arbitrary temperatures and densities. Moreover, our approach correctly takes into account the chemical equilibrium between bound and unbound electrons which is important for the computation of the photoabsorption coefficient and comparisons with experiments.

\section{Numerical results for the ionization potential depression and continuum lowering for hydrogen}\label{s:ipd-hydrogen}
In Sec.~\eqref{sss:ipd}, we derived our main result for the effective ionization potential, $I^{\rm eff,F}_{nl}$, of an arbitrary bound state $|nlm\rangle$ which contains the (positive) ionization potential of an isolated atom, $I_n=|E^0_{n}|$ and a density- and temperature-dependent change of the ionization energy, $\Delta I^{\rm F}_{nl}$. According to Eq.~\eqref{eq:ipd-fermi}, the latter consists of two contributions: the first is $\Delta I_{nl}(n_e,T)$, which arises from the interactions between the particles in the plasma and is negative. The second, $\Delta I^{\rm F}_{nl}(\tilde\chi^*_e)$, is due to quantum exchange effects of fermions and is positive, cf. Fig.~\ref{fig:demo3}.

In most previous treatments, the second term is being neglected, and the change of the ionization potential, $\Delta I_{nl}$, is, therefore, predicted to be negative, ``depressing'' the original ionization potential. As a result, $I^{\rm eff}_{nl}$ eventually vanishes at a critical density, the Mott density, e.g. \cite{bonitz_pop_24}. 
On the other hand, taking into account the Fermi statistics contribution  $\Delta I^{\rm F}_{nl}(\tilde\chi^*_e)$ may change this picture but presently it is unclear how important this effect is.

The key problem of all models and theoretical approaches that are rooted in the chemical picture is the derivation of accurate expressions for the interaction parts of the chemical potentials, $\Delta\mu_a$. This is particularly problematic in parameter ranges, where the degree of ionization changes significantly with density. While accurate results for the interaction between charged particles and between neutral particles exist, the interaction between charged and neutral particles is particularly difficult to describe accurately, which makes chemical models unreliable in the range of the Mott density. 

\subsection{First-principles approach to IPD}
An alternative is to abandon the chemical picture and to use the physical picture that does not discriminate between free and bound electrons. Examples are quantum Monte Carlo, Green functions, and DFT simulations, e.g., \cite{bonitz_pop_24}. This is the strategy we are pursuing in this paper. In the physical approaches also criteria can be established how to identify the fractions of free and bound particles. For quantum Monte Carlo simulations this was discussed in Ref.~\cite{filinov_pre_23}. This allows for a first-principles approach to the IPD that was first proposed in Ref.~\cite{bonitz_pop_24}. Here we continue this analysis providing more details.

\subsubsection{IPD for the ground state}\label{sss:ipd-gs}
To obtain a first result, we assume that no molecules exist and predominantly atoms in the ground state are formed. Our starting point is Eq.~\eqref{eq:pl-function-fermi} which now becomes, using $x_A=1-\alpha$ and Eq.~\eqref{eq:ieff-nl-fermi},
\begin{align}
    \frac{1-\alpha}{\alpha^2} &= \chi_e \, e^{\beta \left(|E_{1s}^0|+\Delta I^{\rm F}_{1s}(\alpha;\chi_e)\right)}\,, \label{eq:ground-state-eq-f}\\
    \Delta I^{\rm F}_{1s}(\alpha;\chi_e) &= \Delta I_0^{\rm F} - \Delta_{1s} 
    \\
    &=\Delta I_{1s}+\Delta I^{\rm F}  \\
    &= 
    \Delta I_0
    + \Delta I^{\rm F}\,.
\end{align}
While in chemical models, the main input in Eq.~\eqref{eq:ground-state-eq-f} are the interaction parts of the chemical potentials and the energy shift of the ground state to yield the degree of ionization, in our first-principles approach we proceed differently:
We assume that the degree of ionization, $\alpha$, is known which allows us to solve for the shift of the ionization energy (IPD), without input of $\Delta \mu$ and $\Delta_{\rm 1s}$:
\begin{align}
    \Delta I^{\rm F}_{1s}(\alpha;\chi_e) 
    &= k_BT \ln\frac{1-\alpha}{\alpha^2\chi_e} - |E^0_{1s}|\,.\label{eq:ipd1}
\end{align}
Finally, if molecules are present, we require as an input also $x_A$, independently of $\alpha$ and need to restore in the numerator of Eq.~\eqref{eq:ground-state-eq-f} $1-\alpha \to x_A$.\\

\subsubsection{IPD for the case of multiple bound states}\label{sss:ipd-many-states}
With temperature increase, contributions of excited states cannot be neglected anymore, and we have to include the full partition function, $\Sigma_A^{\rm PL}$. On the other hand, with increasing density bound states merge into the continuum one by one, starting with the highest excited states. This happens when the energetic position of the continuum, $\Delta I_0^{\rm F}$ crosses the renormalized bound state level, $E_{nl}=E_n^0 - \Delta_{nl}$. In that case the bound state has to be removed from the partition function.
This brings us back to the Saha equation \eqref{eq:pl-function-fermi}, with modification \eqref{eq:ieff-f2} that assures non-negativity of the effective binding energies~\footnote{Due to the finite number of bound states in the presence of the plasma (except for the zero density limit) the canonical partition function is convergent and the Planck-Larkin procedure is not required. However, keeping the latter has the advantage that the partition function is a continuous function of density.},
\begin{align}
    \frac{x_A}{\alpha^2} &=  \chi_e 
\sum_{nl} (2l+1) 
\left\{
e^{\beta I^{\rm eff,F}_{nl}} - 1 - \beta I^{\rm eff,F}_{nl}
\right\} 
\,,\quad\label{eq:pl-kaf}\\
    I^{\rm eff,F}_{nl} &= \mbox{max}\left\{ |E^0_{n}| + \Delta I^{\rm F}_{nl}(n_e,T);0\right\}\,, \label{eq:ieff-f2}\\
    \Delta I^{\rm F}_{nl}(n_e,T) &= \Delta I^{\rm F}_0 - \Delta_{nl}(n_e,T)\,,\\
    \Delta I^{\rm F}_0(n_e,T) &= 
    \Delta I_0(n_e,T)
    + \Delta I^{\rm F}(\tilde\chi_e^*)\,.\label{eq:cont-shift-f}
\end{align}
As discussed in Sec.~\ref{sss:ipd-gs}, we now use known fractions of unbound electrons and atoms, $\alpha$ and $x_A$, and directly solve for the energy position of the continuum, $\Delta I^{\rm F}_0$ [the interaction parts of the chemical potentials are not needed but they can be recovered from this result via Eq.~\eqref{eq:cont-shift}]. In general, this, in addition, requires external input for the level shifts, $\Delta_{nl}$. 

The first approximation, termed ``FPIMC1'' in Ref.~\cite{bonitz_pop_24}, neglects all bound state level shifts, i.e., $\Delta_{nl} \to 0$ and includes only the continuum shift. 
This leads to the simpler set of equations,
\begin{align}
    \frac{x_A}{\alpha^2} &=  \chi_e 
\sum_{n} n^2 
\left\{
e^{\beta I^{\rm eff,F}_{n}} - 1 - \beta I^{\rm eff,F}_{n}
\right\} 
\,,\quad\label{eq:pl-kaf-no-level-shift}\\
    I^{\rm eff,F}_{n} &= \mbox{max}\left\{ |E^0_{n}| + \Delta I^{\rm F}_{0}(n_e,T);0\right\}\,, \label{eq:ieff-f2-no-level-shifts}\\
\Delta_{nl}(n_e,T)\,,\\
    \Delta I^{\rm F}_0(n_e,T) &= 
    \Delta I_0(n_e,T)
    + \Delta I^{\rm F}(\tilde\chi_e^*)\,.\label{eq:cont-shift-f-no-level-shifts}
\end{align}

\subsection{Results for the IPD without level shifts (FPIMC1)}
FPIMC1-results for the continuum, shift, $\Delta I_0^{\rm F}$, are shown in Fig.~\ref{fig:dI_allt}, as a function of the free electron density, $n_e^*$ and $r_s^*=r_s/\alpha^{1/3}$,  for $T=125\,000$K, $T=62\,500$K, $T=31\,250$K, and $T=15\,625$K, by the green triangles. We start by considering the two intermediate temperatures, as they allow us to trace the general trends and properties of our approach.

\subsubsection{FPIMC1-results for the IPD \\ for $T=31\,250$K and $T=62\,500$K}
\begin{figure*}
    \centering
    \includegraphics[width=0.99\textwidth]{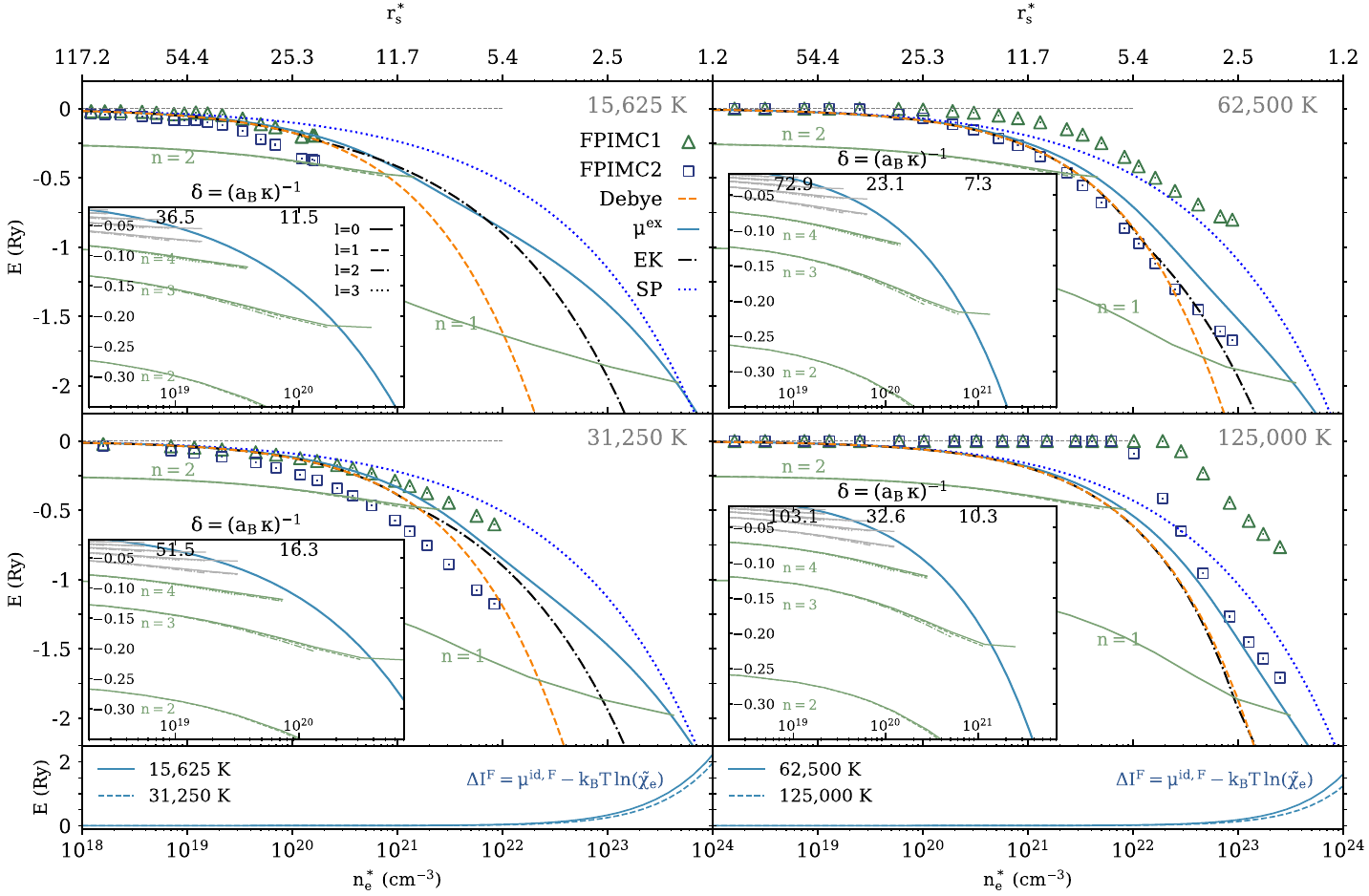}
    \caption{Position of the continuum edge (shift compared to the low-density asymptotic), $\Delta I^F_0$, versus free electron density $n_e^*=\alpha n_e$ (lower x-axis) and $r^*_s$ (upper x-axis), for four temperatures indicated in the plots. Lines with symbols: FPIMC results without (FPIMC1, green triangles) and with (FPIMC2, blue squares) renormalization of the bound state energies. SP and EK denote the models of Stewart and Pyatt and Ecker and Kröll, respectively. $
    \mu^{ex} $ denotes the interaction part of the chemical potentials computed within the chemical model of Ref.~\cite{filinov_pre_23} based on Ebeling's Padé formula \cite{ebeling-richert_85,2017Ebeling}. Bottom row: Fermi statistics contribution, Eq.~\eqref{eq:delta_i-fermi}, to the continuum shift.
    Insets: renormalization of the 2s, 2p, 3s, 3p, 3d, 4s, 4p 4d, and 4f states. Each energy level ends where it  merges in the continuum~\cite{rogers_pra_70}, for a more detailed analysis, see Fig.~\ref{fig:level-ends}. 
}
    \label{fig:dI_allt}
\end{figure*}

Consider first the case of $T=62\,500$K, cf. Fig.~\ref{fig:dI_allt}. Here the position of the continuum goes down monotonically, up to a free electron density of about $n_e^*\approx 10^{23}$cm$^{-3}$, where  $\Delta I_0^{\rm F}$ reaches approximately $-0.75$Ry. At these conditions hydrogen atoms are destabilized by thermal and pressure effects and, at the highest density, the degree of ionization reaches $\alpha \approx 0.8$, cf. Fig.~\ref{fig:ipd_ieff}. For higher densities, fermionic PIMC simulations were not possible due to the fermion sign problem \cite{filinov_pre_23}.
Consider, next, the temperature $T=31\,250$K. Here the overall behavior is very similar. The main difference is the increased stability of atoms, due to reduced thermal excitation. Therefore, at the highest density, the degree of ionization is much lower, $\alpha \approx 0.2$, cf. Fig.~\ref{fig:ipd_ieff}.

Let us now compare these results with existing models. The simplest approximation for the shift of the continuum level is the static Debye approximation (SI units), 
\begin{align}
\Delta I_0|_{\rm Debye}=-\frac{\kappa(n_e^*,T) e^2}{4\pi\epsilon_0}.
\label{eq:debye-shift}    
\end{align}
This result is shown by the dashed yellow lines in Fig.~\ref{fig:dI_allt}. For the inverse screening length $\kappa$ we used the static long wavelength limit of the RPA polarization. This approximation significantly overestimates screening effects, leading to a too large lowering of the continuum. An improved approximation is shown by the solid blue curves in Fig.~\ref{fig:dI_allt}  which are denoted $\mu^{\rm ex}$ and correspond to Ebeling's Padé formulas \cite{ebeling-richert_85,2017Ebeling} for the Coulomb contributions to $\Delta \mu_e+\Delta \mu_p$. This approximation goes beyond the static Debye shift, yielding a reduction of the continuum shift. Evidently, this approximation is significantly closer to our first-principles result. The third approximation depicted in Fig.~\ref{fig:dI_allt} by the dotted blue line (SP) is the Stewart Pyatt result \cite{stewart-pyatt_66}. For the intermediate temperatures, $T=31\,250$K and $T=62\,500$K, this shows the best agreement of all approximations, with the FPIMC1 results. In contrast, the model of Ecker and Kröll (EK, black dash-dotted lines), significantly deviates from the simulations. As expected from its definition, it closely follows the Debye model, for low to moderate densities and elevated temperatures, $T\gtrsim 60\,000$K, whereas, at lower temperatures (stronger coupling), it is closer to the approximation denoted $\mu^{\rm ex}$.

\subsubsection{FPIMC1-results for the IPD  \\for $T=15\,625$K and $T=125\,000$K}
We now turn to the FPIMC1 results at the upper and lower ends of the considered temperature range. Consider first $T=15\,625$K, which corresponds to one tenth of the ground state binding energy of an isolated atom. Thus, one might expect that all electrons are bound in atoms. However, at low density, corresponding to $r_s=100$, the  fraction of bound states is significantly below $100\%$ (the degree of ionization is about $\alpha\approx 0.3$), cf. Fig.~\ref{fig:ipd_ieff}, due to the finite probability of an electron to encounter a proton. When the density increases to about $r_s\approx 10$, practically all electrons are bound in atoms, in the 1s-state, and $\alpha\approx 0$ and remains at this level up to, at least, $r_s=4$. At larger densities one expects an increase of $\alpha$, due to pressure ionization, i.e., IPD. However the density where this sets in could not be resolved because, no PIMC simulations were possible, due to the fermion sign problem. Nevertheless, the PIMC simulations allow us to compute the IPD and continuum lowering also in the low density range, $r_s\gtrsim 4$, cf. Fig.~\ref{fig:dI_allt}. Due to the low $\alpha$-values, the maximum free electron density is rather low and equals approximately $n_e^* \approx 2 \cdot10^{20}$cm$^{-3}$ and, at this point, the continuum lowering reaches about $0.25$Ry. At these low electron densities there is close agreement of most models with the PIMC results. Only the Stewart Pyatt model deviates significantly, underestimating the continuum lowering at this temperature.

Let us now consider the maximum temperature, $T=125\,000$K, which corresponds to about $0.8$Ry. Here the degree of ionization is always high, not falling below $\alpha \approx 0.75$ [cf. Fig.~\ref{fig:ipd_ieff}], whereas the bound electrons are distributed over many bound states, from the ground state to high excited states. 
%
Interestingly, the value of $\Delta I_0^{\rm F}$ that fulfills Eq.~\eqref{eq:pl-function-fermi} remains close to zero for a broad range of low densities. This is caused by the combination of two effects: first, for higher temperatures, the Planck-Larkin partition function decays weakly with increasingly negative values of $\Delta I_0^{\rm F}$ and, second, a large value of the l.h.s. of Eq.~\eqref{eq:pl-function-fermi}, i.e., of $\frac{x_A}{\alpha^2}$, as well as of$ \frac{1}{\chi_e}$, for low densities.
For this reason, the position of the continuum remains close to zero, for densities up to about $n_e^*\approx 2\cdot 10^{22}$cm$^{-3}$, where $\alpha$ reaches its minimum, cf. Fig.~\ref{fig:ipd_ieff}. Beyond this point, the continuum edge quickly decreases and $\alpha$ rapidly increases. The observed almost step-like behavior of the continuum edge, at this high temperature, is unexpected and is qualitatively different from all models. However, none of the models takes into account excited atomic bound states.

To summarize, we have presented first-principles results for the continuum lowering that are based on PIMC-simulation data for the degree of ionization. While these results take into account the effect of the plasma medium on the continuum position, they, however, neglect the influence of the plasma on the bound states. This effect is taken into account, approximately, in our second approach, termed ``FPIMC2''~\cite{bonitz_pop_24}.

\subsection{Results for the IPD with level shifts\\ taken into account (FPIMC2)}
As we illustrated in  Fig.~\ref{fig:demo1}, 
the plasma-induced shifts of bound states  are expected to be negative. This means that, for a given value of $\alpha$ and for a given density, where bound states merge with the continuum, the latter has to decrease more strongly than in the case of zero level shifts. This is indeed what we will observe below.

\subsubsection{Energy shifts of the lowest levels}\label{sss:level-shifts}
The key question in  ``FPIMC2'' is how the level shifts can be computed. While PIMC simulations yield an effective fraction of atoms, $x_A$, they do not provide information about the occupancy of various bound states and on their effective binding energies. Therefore, one has to reconsider the bound state problem of an electron and a proton, but now under conditions where the pair is embedded into a partially ionized hydrogen plasma. This is a formidable many-body problem for which only approximate solutions exist. The most advanced approach is based on the Bethe-Salpeter equation (BSE) for the electron-proton Green function (or density matrix) developed by Zimmermann \textit{et al.} \cite{zimmermann_pssb_78} and Haug et al.~\cite{haug_pqe_84}. The main effects that are of importance are static and dynamic screening of the electron-proton attraction, strong coupling, quantum and spin effects in the plasma, including Pauli blocking, and the interaction of electrons and protons with atoms and molecules. The solution of the BSE has been discussed in many papers, e.g. \cite{seidel_pre_95} and \cite{roepke_pre_19} and text books \cite{green-book,blue-book}.
Presently, a selfconsistent account of all the effects mentioned above has not been possible. We, therefore, will resort here to a simpler approach that takes into account only static screening of the electron-proton attraction,  which is expected to be the dominant effect at low densities.

In fact, the hydrogen Schrödinger equation, with the Coulomb potential replaced by a Debye potential, 
\begin{align}
V_D(r;\kappa):=\frac{e^2}{4\pi\epsilon_0}    \frac{e^{-\kappa r}}{r}\,,
\label{eq:debye-pot}
\end{align}
has been solved already by Rogers \textit{et al.}~\cite{rogers_pra_70}, for different values of the dimensionless parameter $\delta=(a_B\kappa)^{-1}$ -- the ratio of the screening length $\kappa^{-1}$ to the Bohr radius. More recently Onate \textit{et al.}~\cite{onate_jtap_16} presented an alternative approach, and their results were used in Ref.~\cite{bonitz_pop_24}. Due to some inconsistencies in the data of Ref.~\cite{onate_jtap_16}, here we will employ the level shifts reported in Ref.~\cite{rogers_pra_70}. The final step is to relate the screening parameter $\kappa$ to the plasma density, $n_e^*$ and temperature $T$. This is done by using the static long wavelength limit of the RPA polarization which interpolates between the Debye radius, at low degeneracy, and the Thomas-Fermi length, at high degeneracy. In addition, Ebeling’s Padé formula is applied to the screening length to account for higher order screening effects, cf. curves denoted $\mu^{ex}$. 
The results for the level shifts for the principal quantum numbers, $n=2 \dots 9$, and four temperatures are depicted in the insets of Fig.~\ref{fig:dI_allt}. In addition, the first two levels,  $n=1$ and $n=2$, are also shown in the main plots, for the entire density range. Note that the screening removes the $l$-degeneracy, so results for different $l$ are shown. The level energies are plotted up to the density where they merge into the continuum (i.e., the eigenstate vanishes), as reported in Ref.~\cite{rogers_pra_70}. 

\begin{figure}
    \centering
    \includegraphics[width=0.49\textwidth]{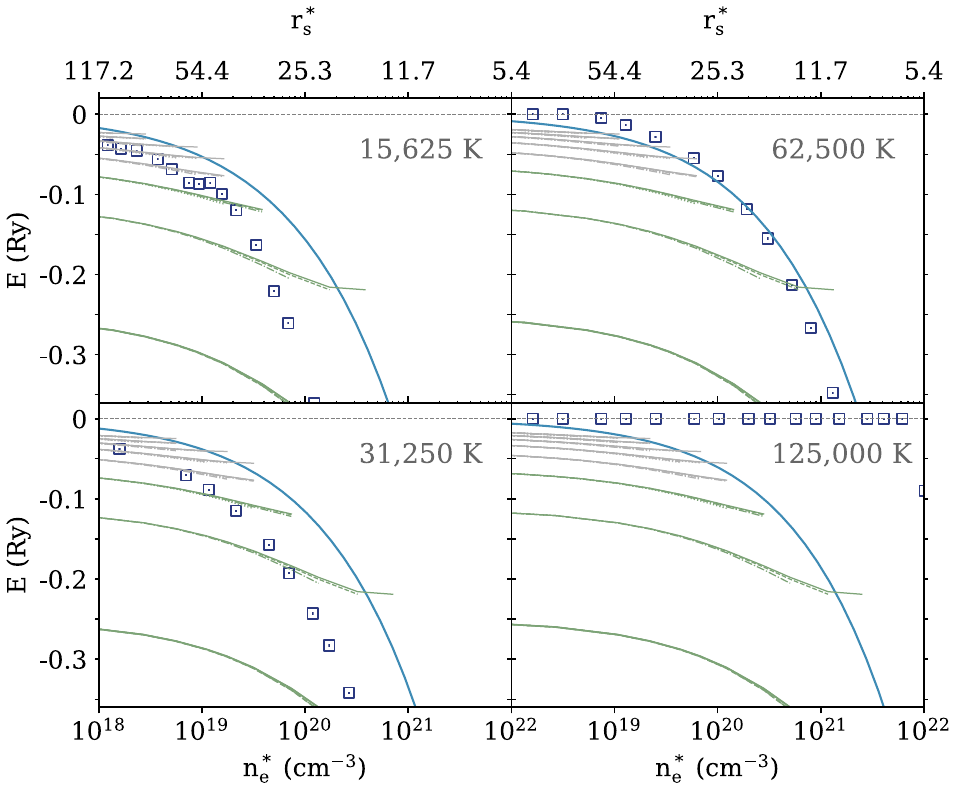}
    \caption{Renormalized energies of excited states (zoom into insets of Fig.~\ref{fig:dI_allt}), according to Ref.~\cite{rogers_pra_70} (full green lines) for principal quantum numbers $n = 2 \dots 9$ and position of the continuum according to the FPIMC2-approach (squares), for four temperatures indicated in the figures. The blue line is the continuum position according to Ebeling's Padé formula \cite{ebeling-richert_85,2017Ebeling} (lines $\mu^{\rm ex}$ in Fig.~\ref{fig:dI_allt}).
}
    \label{fig:level-ends}
\end{figure}

It is interesting to compare this to the points where the bound states cross  the FPIMC2-result for the continuum (blue squares) which we regard as more accurate than the solution of the Schrödinger equation. For a detailed comparison, we replot the curves for the excited states with higher resolution in Fig.~\ref{fig:level-ends}. 
 Let us look separately at the four temperatures:
\begin{description}
    \item[T=15\,625 K] All bound state levels extend far beyond the FPIMC2-continuum. Instead, they terminate close to the blue Padé curve.
    \item[T=31\,250 K] The behavior is similar as for the lower temperature, but the n=2 level terminates closer to the FPIMC2-continuum.
    \item[T=62\,500 K] The agreement with the FPIMC2-data improves further, for all levels.
    \item[T=125\,000 K] Here the ground state ends almost exactly at the FPIMC2-continuum, whereas the excited state levels terminate earlier.   
\end{description}
Let us briefly discuss the implications of this comparison for our FPIMC2-approach.
Combining the renormalized bound state levels of Ref.~\cite{rogers_pra_70} with the PIMC data for the degree of ionization is not entirely consistent. Therefore, we can turn this into a consistency check: the present FPIMC2-procedure, in combination with the data of Ref.~\cite{rogers_pra_70}, is best validated in cases where the levels end in the vicinity of the FPIMC2-continuum.
Thus, the best agreement is observed for the two highest temperatures, in particular, for $T=62\,500$K. For lower temperatures, only the low-density part of the bound state energies are contributing to the FPIMC2-solution. But, at low densities, the static screening approximation can be regarded accurate, which again supports the FPIMC2-results.\\

\subsubsection{FPIMC2-results for the continuum position, $\Delta I_0^{\rm F}$, for four temperatures}\label{sss:continuum-fpimc2}
\begin{figure}
    \centering
    \includegraphics[width=0.49\textwidth]{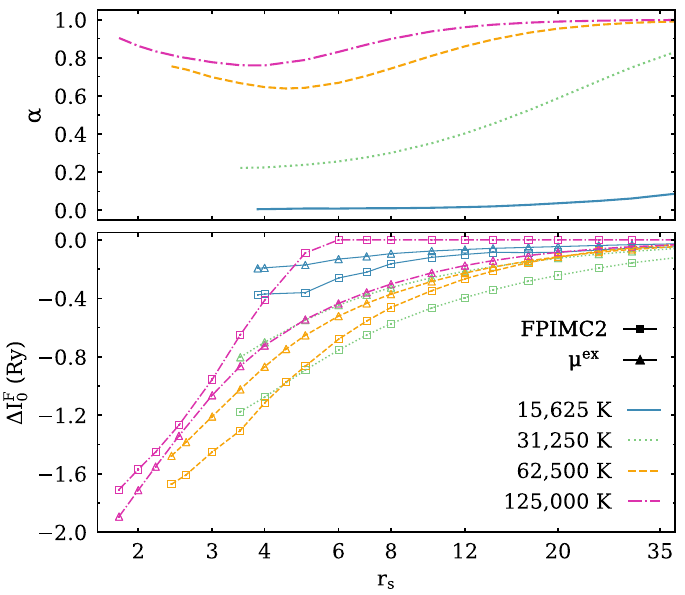}
    \caption{\textbf{Top:} Total density dependence of the
     free electron fractions (degree of ionization). \textbf{Bottom:} Continuum position (shift) from FPIMC2-simulations, vs. Padé formula ($\mu^{ex}$), for four temperatures.
    }
    \label{fig:deltai-allt}
\end{figure}

We now use the level shifts of Sec.~\ref{sss:level-shifts} as input to the full set of equations \eqref{eq:pl-kaf} and \eqref{eq:ieff-f2} for computation of the continuum shift, $\Delta I_0^{\rm F}$. 
 The numerical results are presented in Fig.~\ref{fig:dI_allt} by the blue squares. 
As expected, the results for $\Delta I_0^{\rm F}$ are below the FPIMC1-results, where level shifts are neglected (green triangles). This is observed for all temperatures, and the differences increase monotonically with the free electron density. 
This trend is mainly caused by the density dependence of the ground state level. For free electron densities above $10^{22} \text{cm}^{-3}$, excited states already vanished into the continuum, so at these densities, there is no influence of higher energy levels. However, even for lower densites, $(10^{18} - 10^{22}) \text{cm}^{-3}$ their effect is small: artificially excluding  higher energy levels from the partition function decreases the absolute value of the FPIMC2-result only 0.05 Ry, at most. This maximum deviation occurs at $T= 62\,500$K, around $10^{20} \text{cm}^{-3}$ and is much smaller for the other temperatures and densities, where the IPD result is not zero. Therefore, the shape of the FPIMC2 curve is mainly determined by the shape of the ground state energy level, aside from the FPIMC input data, of course. For the lower temperatures, the ground state energy of Ref.~\cite{rogers_pra_70} starts to decrease already for smaller densities than in the case of larger temperatures. This is the origin of the comparably large absolute value of the FPIMC2 data points at $31,250$ K, for low densities, cf. also Fig.~\ref{fig:deltai-allt}.  \\

 It is interesting to compare our first-principles results to those of the other models. At low temperatures, $T\lesssim 30\,000$K, the FPIMC2-results are significantly below all other results, even below the Debye approximation \eqref{eq:debye-shift}, cf. Fig.~\ref{fig:dI_allt}. For $T= 62\,500$K, the FPIMC2-data are rather close to the Debye and Ecker-Kröll models, whereas for $T=125\,000$K these two models overestimate the continuum shift  compared to the FPIMC2-result. The Stewart Pyatt model, overall, deviates more strongly from the FPIMC2-data than the other models.

In Fig.~\ref{fig:deltai-allt}, the FPIMC2-results of Fig.~\ref{fig:dI_allt} are replotted as a function of the total density parameter $r_s$, which is related to the degree of ionization $\alpha$ and the free electron density (used in Fig.~\ref{fig:dI_allt}) via $n_e^* = \frac{3}{4\pi} \left ( r_s a_B \right)^{-3} \alpha $.\\

\subsection{Effective binding energy of the 1s-state, $I_1^{\rm eff}$: FPIMC1- and FPIMC2-results}
 
\begin{figure}
    \centering
    \includegraphics[width=0.49\textwidth]{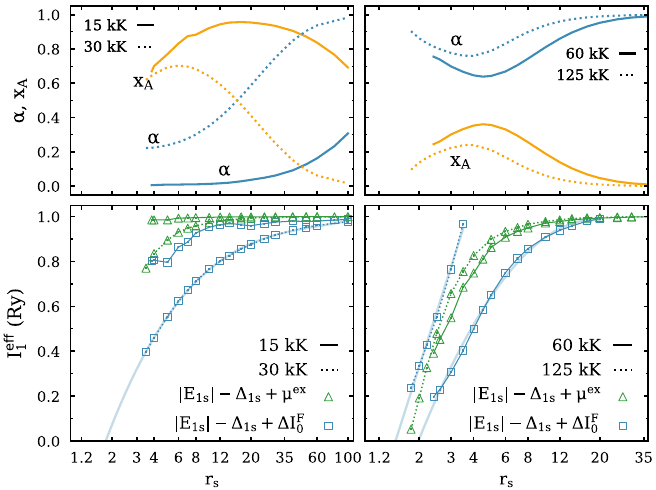}
    \caption{\textbf{Top:} Density dependence of
     free electron and atom fractions calculated with FPIMC-simulations in Ref.~\cite{filinov_pre_23}. The molecular fraction that is also present at the two lower temperatures (therefore, $\alpha+x_A < 1$) is not shown. \textbf{Bottom:} Effective ionization energy of the  ground state, $I^{\rm eff}_1$, with level shift $\Delta_{1s}$ included, for $31\,250$ K and $15\,625$ K (left), and  $62\,500$ K and $125\,000$ K (right). 
     Triangles: continuum shift based on Ebeling's Padé formula. Squares: continuum shift from FPIMC2-data. Light blue lines: analytical fit and extrapolation to the Mott density, for details, see text.
}
    \label{fig:ipd_ieff}
\end{figure}

 Finally, we compute the effective binding energy of the ground state, $I_1^{\rm eff}$, i.e. the energy distance from the renormalized ground state level, $E_{1}$, to the renormalized continuum edge, $\Delta I_0^{\rm F}$. While none of the two quantities is directly measurable, their difference, $I_1^{\rm eff}$, is and, therefore, plays a central role for comparison with experiments, cf. Sec.~\ref{s:dis}. The results of our first-principles simulations for $I_1^{\rm eff}$ are shown  in Fig.~\ref{fig:ipd_ieff}. In the top row we plot the PIMC-results for the degree of ionization which are crucial to transform the previous results, that were plotted vs. the free electron density and $r_s^*$, via the total density parameter $r_s=\alpha^{1/3} \, r_s^*$.\\

In the bottom row, the effective ionization energy $I_1^{\text{eff}} = |E_{\text{1s}}| - \Delta_{\text{1s}} + \Delta I_0^{\rm F} $ is shown where $|E_{\text{1s}}| - \Delta_{\text{1s}}$ is the absolute value of the renormalized ground state energy from Ref.~\cite{rogers_pra_70}. For the continuum lowering, $\Delta I_0^{\rm F}$, the FPIMC2-values were used (blue squares) that were already shown in Fig.~\ref{fig:dI_allt} (squares). Note that the FPIMC2-results and FPIMC1-results for $I_1^{\text{eff}}$ are practically identical.
The reason is that, when solving
Eq.~\eqref{eq:pl-function-fermi} for the continuum lowering, $\Delta I_0^{\rm F}$, the r.h.s. of the equation $\Sigma_A^{\text{PL}} \left( \Delta I_0^{\rm F} \right) = \frac{x_A}{\alpha^2} \frac{1}{\chi_e}$, is known from PIMC-simulations, and the equality with the l.h.s. is achieved by varying $\Delta I_0^{\rm F}$. If the partition function contains the renormalized energy levels (FPIMC2) this leads to an increased value of $\Delta I_0^{\rm F}$ that compensates the bound state energy shift. As long as the ground state dominates the partition function (as is the case here), the compensation is practically exact.
Therefore, we do not plot the FPIMC1-data. Instead, as a second data set for the effective ionization energy (triangles), in addition to the FPIMC2-results (squares), we included in Fig.~\ref{fig:ipd_ieff} data that are derived from the continuum shift that uses $\mu^{ex}$ (Ebeling’s Padé formula).  

Let us now discuss the behavior of $I_1^{\text{eff}}$  associated with $\mu^{ex}$ (green triangles). It measures the energy distance from the ground state to the continuum curve $\mu^{ex}$ and reflects the behavior common to chemical models. 
For all 4 temperatures, $I_1^{\text{eff}}$ starts at and remains close to 1 Ry for low densities, corresponding to $r_s \approx  100 \dots 10$.
 For $r_s \lesssim$ 10, $I_1^{\text{eff}}$ begins to decrease because the continuum lowering is faster than that of the ground state. An exception is the lowest temperature, $T\sim 15\,000$ K, where $I_1^{\text{eff}}$ does not decline up to at least $r_s=4$. Due to the low degree of ionization, the free electron density is too small to produce a significant shift of the continuum. With increasing temperature (at a fixed value of $r_s$), the decrease of $I_1^{\text{eff}}$ proceeds faster because the degree of ionization and the free electron densities are growing. This trend continues up to about $100\,000$ K after which it is reversed: for $T=125\,000$ K,  $I_1^{\text{eff}}$ is larger than for $T = 62\,500$ K. This is explained by the non-monotonic temperature dependence of $\mu^{ex}$, cf. Fig.~\ref{fig:deltai-allt}. 

Consider now $I_1^{\text{eff}}$ calculated from the FPIMC2-data (squares). For $15\,625$ K and $31\,250$ K, the FPIMC2-data points are always below the $\mu^{ex}$-data, reflecting the behavior of the continuum position of the two approaches, cf.~Fig.~\ref{fig:deltai-allt}. 
For $62\,500$ K the same trend is observed, but the curves are closer to one another, except for very low densities, where the FPIMC2 data are located slightly above $\mu^{ex}$ (Fig.~\ref{fig:level-ends}). 
At $125\,000$ K, the FPIMC2-results (squares) are above the Padé approximation (triangles), for all densities. This is due to the different behaviors of the continuum position for the two approaches, cf. Figs.~\ref{fig:level-ends} and 
~\ref{fig:dI_allt}). Particularly striking is the rapid decay of the FPIMC2-binding energy, for $r_s\lesssim 3.5$, which is caused by the density dependence of the continuum shift, cf. Fig.~\ref{fig:level-ends}.
For lower densities we observe  $I_1^{\text{eff}} $ values slightly larger than one (not shown), which is attributed to inconsistencies of the FPIMC-input data and the level shifts.

\subsection{Mott density}
Finally, note that the symbols in Fig.~\ref{fig:ipd_ieff} stop at a finite value of $I^{\rm eff}_1$ and do not reach zero. The last point corresponds to a minimum value of $r_s$, in the range of $2\dots 4$, depending on temperature, due to the fermion sign problem. At the same time, it is tempting to extrapolate the curves towards the Mott density, $r^{\rm Mott}_s(T)$, i.e., the value where $I_1^{\text{eff}}(T)\to 0$. 
We used the easiest method for an extrapolation that incorporates information about all existing data points -- a simple polynomial fit. At $30\,000$ K, a third order polynomial was applied, at $60\,000$ K a quadratic function, and at $125\,000$ K a simple linear fit. These fits are included in Fig.~\ref{fig:ipd_ieff} as additional light blue lines. It is important to place emphasis on the fact that the underlying model is not known, i.e., the actual shape of $I_1^{\text{eff}}(r_s)$, from the last data point to the Mott-density may, in principle, deviate from the previous trend. 
Hence, it is not possible to reliably estimate the uncertainties from the function fits. Instead, the results for the Mott-density from the extrapolation are identified in the form of a confidence interval, based on reasonable ways to extrapolate the data points in a smooth way.

The extrapolation 
yields the following results: $r^{\rm Mott}_s(30\,000)\approx 1.35\dots 1.8$, $r^{\rm Mott}_s(60\,000)\approx 1.5\dots 2.0$ and $r^{\rm Mott}_s(125\,000)\approx 1.5$. For $T=15\,000$ K no extrapolation is possible. 
On physical grounds we expect that $r^{\rm Mott}_s(T)$ increases with temperature, as is the case for the two lowest temperatures.  The result for $T=125\,000$ K, however, deviates from this trend, as does the overall behavior of the curve $I_1^{\text{eff}}(r_s)$, in this case. The reason for this behavior remains to be explored. Interestingly, the obtained values  for the Mott density differ slightly from previous estimates, including the value
$r_s \approx 1.2$  reported in Ref.~\cite{bonitz_prl_5}. We return to this issue in Sec.~\ref{s:dis}.

\subsection{Sensitivity of the results to the criteria for free and bound states}\label{ss:sensitivity}
Our procedure to compute the continuum shift and the effective binding energy based on first-principle PIMC simulations that are based on the ``physical picture'' where no sharp distinction between free and bound electrons exists [for a discussion, see Ref.~\cite{bonitz_pop_24}]. Thus, in order to identify the fractions of free electrons and electrons bound in atoms and molecules, respectively, formal criteria have to be introduced. Thus, it may seem that the results carry some degree of arbitrariness. However, this is not the case because it is straightforward to test the sensitivity of our results on the chosen criterion.
The procedure to identify free electrons and bound states that was applied to the FPIMC-data was explained in Ref.~\cite{filinov_pre_23} and will not be reproduced here. In that paper also the influence of the criterion to distinguish atoms and molecules was investigated. In particular, when the critical proton-proton distance was changed in reasonable limits (about $15\%$), this gave rise to changes of $x_A$ by not more than $3\%$. We therefore re-analyzed the results when $x_A$ was artificially varied by up to $10\%$ and observed that $I_1^{\text{eff}}(r_s)$ changes by not more than $2\%$. Similarly, if $\alpha$ is varied,  $I_1^{\text{eff}}(r_s)$ changes approximately twice as much. 

\begin{figure}
    \centering
    \includegraphics[width=0.98\linewidth]{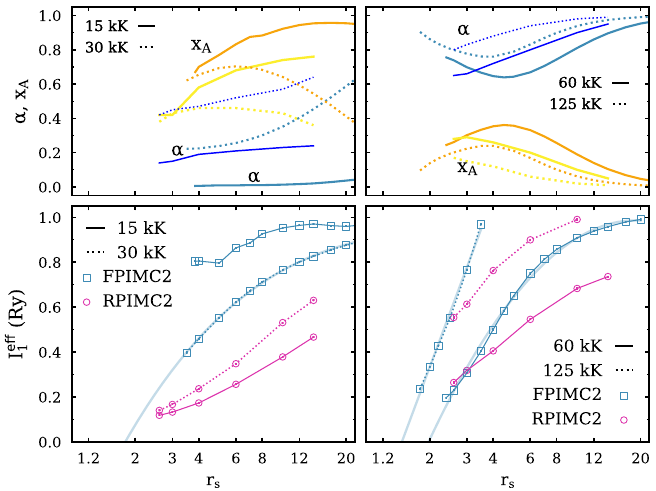}
    \caption{Comparison of the present FPIMC2-results (from Fig.~\ref{fig:ipd_ieff}) to RPIMC-data of Ref.~\cite{Militzer_PRE_2001} for the effective ground state binding energy. The data for $\alpha$ (top panels) are significantly higher in RPIMC than in FPIMC. This leads to a shift of the IPD result toward larger values of the free electron density and a larger absolute value of the IPD,  due to a decrease of the l.h.s. of Eq.~(\ref{eq:pl-function-fermi}). The combination of both effects results in a decreased value of the binding energy $I_1^{\rm eff}$ (bottom panels), especially at lower temperatures. Note that the x-axis extends over the density range where RPIMC data are available ($r_s\le 12$).}
    \label{fig:rpimc-comparison}
\end{figure}

As an additional test of the procedure  we used the RPIMC-results of Militzer and Ceperley for the degree of ionization and for the atom fraction of hydrogen, in the same density-temperature range \cite{Militzer_PRE_2001}. Use of the RPIMC-data instead of the FPIMC-results, affects the IPD obtained by our approach much more strongly than the variation of the criterion for $\alpha$ and $x_A$, as explained above. The largest difference between both data sets is observed at the lowest temperature, $T=15\,625$ K, where RPIMC yields $\alpha$ ($x_A$)-values that are about $0.2$ larger (smaller) than FPIMC. This has a dramatic effect on the IPD, as we show in Fig.~\ref{fig:rpimc-comparison}, where we include the comparison of both data sets for procedure 2 (with finite renormalization of the bound states).
Also, the RPIMC-curves exhibit an unexpected softening at high densities that would lead to a too high Mott density, which has to be attributed to the nodal errors of the method.

We conclude that the presented analysis provides strong support for the validity of our procedure, in combination with FPIMC-data.

\section{Towards estimates of IPD for $Z$-fold charged ions. Influence of the Fermi statistics}\label{s:ipd-zfold}
The influence of the Fermi statistics on the Saha equation led to the Fermi barrier, $\Delta I^{\rm F}$, that tends to stabilize bound states. This quantity has been plotted, for comparison, in the bottom panels of Fig.~\ref{fig:dI_allt}. At the highest densities for which FPIMC data are available, $\Delta I^{\rm F}$ did not exceed $1\dots 2$eV but in the range of the Mott point, $r_s\sim 1.2$, $\Delta I^{\rm F}$ is on the order of $1\dots 2$ Rydberg.  

Since the Fermi barrier is, at high degeneracy, increasing with density approximately as the Fermi energy, $\sim n_e^{*2/3}$, we expect its role to be more pronounced in the ionization of high-$Z$ atoms. To further analyze this effect we extend our previous analysis to hydrogen-like bound states that are composed of a single electron and a $Z$-fold charged nucleus. 
Ionization of the ground state is then commonly known as K-shell ionization.
Compared to the case of hydrogen, 
the Rydberg energy and Bohr radius change according to $E^H_R \to E_R^Z = Z^2 E_R^H$ and $a^H_B \to a_B^Z =  a_B^H/Z$. 
Assuming, for an estimate, that the Mott density still corresponds to $r^Z_s \approx 1.2$, where $r_s^Z=a^Z/a_B^Z$, with $a^Z$ denoting the mean distance between electrons in this plasma, this then corresponds to an electronic Mott density that is $Z^3$ times higher than for hydrogen \cite{bonitz_prl_5}. Consequently, close to the Mott point, electron degeneracy plays a much larger role than for hydrogen, and the Fermi shift, $\Delta I^{\rm F}$, is expected to have a larger impact on the ioinzation equilibrium and the IPD.

\subsection{Modification of the Saha equation}\label{sss:saha-z}
We now discuss how the Saha equation \eqref{eq:ground-state-eq-f} is modified. 
Again, we assume that no molecules, but only $Z$-fold charged and $Z$-1-fold charged ions exist. For simplification we assume that predominantly ions in the ground state are present. 
We will denote the density of $Z$-fold charged ions by $n_{Z}$ and that of $Z$-1 fold charged ions by $n_{Z-1}$, and the total ion density is $n_i=n_{Z}+n_{Z-1}$. The number fraction of bare nuclei is then $\alpha=n_Z/n_i$.  Charge neutrality requires that the free electron density is
\begin{align}
    n_e^* &= Z n_{Z} + (Z-1)n_{Z-1} = \nonumber\\
          &= n_i (Z-1 + \alpha) = \bar Z n_i\,, \label{eq:nestar-z}\\
    \bar Z &= Z \alpha + (Z-1)(1-\alpha) = Z-1+\alpha\,,\label{eq:zbar}
\end{align}
where we also introduced the average ionic charge, $\bar Z$.
The total electron density contains also $n_{Z-1}$ electrons bound in the $Z$-1-fold charged ions, i.e., 
\begin{align}
n_e = n_e^* + n_{Z-1} = Z n_i\,. \label{eq:nez}    
\end{align}

Starting again with the classical case, our previous equations,  \eqref{eq:saha-statej}, are modified by the appearance of general statistical factors,
\begin{align}
    \frac{n_{Z-1}}{n_e^* n_Z}\bigg|_{\rm class} &=
    \frac{\Lambda_e^3}{2}\frac{g_Z}{g_{Z-1}}e^{\beta(E_1^{Z}-E_1^{Z-1})} e^{\beta \Delta I_0^{Z-1}}\,,\\
    \Delta I_0^{Z-1} &= \Delta \mu_e +  \Delta \mu_{Z} - \Delta \mu_{Z-1}\,,
\end{align}
where $E_1^{Z-1}$ is the lowest binding energy of an electron in the $Z$-1-fold charged ion in the plasma (i.e., including the level shift), whereas $E_1^{Z}=0$ is the corresponding ``binding energy'' of the bare nucleus.
We bring this equation to the previous dimensionless form, by multiplying by $n_e$, Eq.~\eqref{eq:nez}, 
\begin{align}
    \frac{(1-\alpha) Z}{\alpha(Z-1+\alpha)}\bigg|_{\rm class} &= \frac{\chi_e}{2}\frac{g_Z}{g_{Z-1}} \, 
    e^{\beta \left(|E_{1s}^{Z-1}|+\Delta I_0^{Z-1}\right)}\,. \label{eq:ground-state-eq-z}
\end{align}
Finally, we restore the spin statistics of the electrons, performing exactly the same steps as for hydrogen and obtain
\begin{align}
    \frac{(1-\alpha) Z}{\alpha (Z-1+\alpha)} &= \frac{\chi_e}{2}\frac{g_Z}{g_{Z-1}} \, 
    e^{\beta \left(|E_{1s}^{Z-1}|+\Delta I_0^{Z-1}+\Delta I^{\rm F}\right)}\,, \label{eq:ground-state-eq-zfermi}
\end{align}
where the continuum shift is now given by the sum $\Delta I_0^{\rm F, Z-1}(n_e, T; \alpha, Z) = \Delta I_0^{Z-1}+\Delta I^{\rm F}$. If we would have access to the $Z$-fold charged ion fraction $\alpha$ [or to $\bar Z$, Eq.~\eqref{eq:zbar}], e.g. via experiment or simulation, we could directly compute the continuum shift and, from it and the level shift, the ionization potential depression, following the approaches developed in Sec.~\ref{s:ipd-hydrogen}.
The value of the Fermi shift, at high degeneracy, is estimated in Sec.~\ref{sss:fermi-shift-highn}.

Let us briefly discuss the degeneracy factors, for the example of beryllium. Here $n_Z$ corresponds to a fourfold charged nucleus. If its quantum effects can be neglected, we can choose $g_4 \to 1$. Correspondingly, the degeneracy of the bound state, i.e., of Be$^{3+}$, is $g_3=g_e g_4 \to  2$. 

\subsection{Estimate of screening effects on IPD at high densities}
So far no rigorous PIMC-data are available for the degree of ionization of high-Z plasmas. Therefore, here we provided a simple estimate for the IPD and the influence of the Fermi statistics shift, by using a Debye-type screening model. Based on the hydrogen results [cf. Fig.~\ref{fig:dI_allt}] this, most likeley, overestimates the screening.

For arbitrary $Z$, the ionic contribution becomes
\begin{align}
    \kappa^2_i &= \beta  \frac{e_0^2}{\epsilon_0} \left\{  Z^2 n_Z + (Z-1)^2n_{Z-1}\right\}\,,
    \label{eq:kappai-debyez}
\end{align}
and the square of the total (electron plus ion) inverse screening length, in the classical limit, is 
\begin{align}
    \kappa^2|_{\rm class} &= \beta\frac{e_0^2}{\epsilon_0}n_e (\bar Z+\alpha)\,,
\end{align}
where we used Eq.~\eqref{eq:zbar}.
On the other hand, with quantum electrons, the electronic screening parameter changes to \cite{blue-book}
\begin{align}
    \kappa_e^2 &=\frac{e_0^2}{\epsilon_0}\,\frac{g_e}{\Lambda_e^3}\frac{d}{d\mu}I_{1/2}(\beta\mu_e) = \\
    &=\frac{e_0^2}{\epsilon_0}\beta \,\frac{g_e}{\Lambda_e^3}I_{-1/2}(\beta\mu_e)\,.
\end{align}
In the following, we concentrate on strong electron degeneracy, $\tilde \chi_e \gtrsim 5.5$, for which the Fermi integral can be approximated as \cite{blue-book}
\begin{align}
    I_{-1/2}(\beta\mu) &= \frac{\tilde\chi_e^{1/3}}{0.806 + 0.4535 \, \tilde\chi_e^{-4/3}+ 1.7 \, \tilde\chi_e^{-8/3}} \approx \\
    &\approx 1.241 \tilde\chi_e^{1/3}\left\{1 - 0.563 \tilde\chi_e^{-4/3} \right\}\,,\label{eq:i-expansion-high-n}
\end{align}
and the screening parameter becomes
\begin{align}
    \kappa^2 =&\kappa_e^2+\kappa_i^2 =\\
    =&\beta\frac{e_0^2}{\epsilon_0}\left\{Z^2 n_Z + (Z-1)^2n_{Z-1} + n_e\tilde \chi_e^{-1}  I_{-1/2}(\beta\mu) \right\} = \\
    =&\beta n_e\frac{e_0^2}{\epsilon_0}\bigg\{Z \alpha + (1-\alpha)\frac{(Z-1)^2}{Z} + \\
    & + 1.241 \,\tilde \chi_e^{-2/3} \left[ 1 - 0.563 \tilde\chi_e^{-4/3}\right]  \bigg\}\,.\label{eq:kappa2-degenerate}
\end{align}
This result is clearly dominated by the ionic contribution, in particular, for strong degeneracy and $Z>1$. Using the Debye approximation \eqref{eq:debye-shift} for the continuum shift, we obtain 
\begin{align}
\Delta I_0|_{\rm Debye}=-\frac{1}{2}\frac{\kappa(n_e^*,T) e_0^2}{4\pi\epsilon_0}\left\{ 1 + \bar Z^2 \right\}\,,
\label{eq:debye-shift-z}    
\end{align}    
where the first term is due to the electrons and the second due to the ions and $\kappa$ is given by Eq.~\eqref{eq:kappa2-degenerate}.
In the case of strong ionic correlations, $\Gamma_i>1$, corresponding approximately to $\kappa \,n_i^{-1/3}\gtrsim 1$, a better approximation is given by the ion sphere model.

\subsection{High-density approximation for the \\Fermi barrier}\label{sss:fermi-shift-highn}
Using the high-degeneracy expansion of the Fermi integral we obtain \cite{blue-book}
\begin{align}
    \Delta I^{\rm F} & \approx 1.21\, k_BT  \tilde\chi_e^{2/3}\left( 1 - \frac{0.562}{\tilde\chi_e^{4/3}}  - \frac{\ln{\tilde\chi_e}}{1.21 \tilde\chi_e^{2/3}} \right).\; 
    \label{eq:fermi-shift-high-n}
\end{align}
The first term contains the ground state Fermi energy and is independent of temperature, whereas the next describes the finite temperature corrections (generalizing the Sommerfeld expansion). Note that the last term, accounting for the classical limit, is significant and cannot be neglected. The leading order density dependence is that of the Fermi energy, $\Delta I^{\rm F} \sim n_e^{2/3}$.

\subsection{Influence of the Fermi statistics on IPD}
We now extend our hydrogen result for the effective ionization energy, Eq.~\eqref{eq:ipd-fermi}, to the case of $Z$-fold charged ions:
\begin{align}
    I^{\rm eff,F}_{nl}(n_e,T) = &  |E_n^0| - \Delta_{nl}(n_e,T) + 
    \nonumber\\
    &+ \Delta I_{0}(n_e,T)+ \Delta I^{\rm F}(n_e,T)\,.
    \label{eq:ieff-nl-fermi-z}  
\end{align}
While the Fermi barrier is known from Eq.~\eqref{eq:fermi-shift-high-n}, the shift of the continuum, $\Delta I_0$, and of the bound states, $\Delta_{nl}$, are not. The Debye approximation for the continuum shift, Eq.~\eqref{eq:debye-shift-z},
suggests a scaling $\Delta I_0|_{\rm Debye} \sim (Z^3 n_e \beta)^{1/2}$ which means that one would expect the relative importance of the Fermi barrier, $\Delta I^{\rm F}$, to grow with density. However, this neglects several additional effects. First, with increasing density, the continuum shift will not only be affected by screening, but also by exchange, i.e., the Fock selfenergy, which is negative and increases the continuum lowering. Second, exchange effects and Pauli blocking also affect the bound state energies, reducing the electron binding energy. In particular, it is clear that, if the electron gas is so dense that $E_F \ge E_R^Z$, no bound states can exist. Thus, exchange gives rise to two competing effects: a lowering of the binding energy and the Fermi barrier effect, to which Coulomb correlation effects, mostly due to the ions, have to be added. Therefore, a theory of IPD in high-Z plasmas has to selfconsistently take into account all these effects.

As was discussed in Sec.~\ref{s:intro}, the most common approximations -- the Ecker-Kröll \cite{ecker-kroell_63} and Stewart-Pyatt \cite{stewart-pyatt_66} models do not account for the level shifts and provide only an incomplete treatment of the many-body and quantum effects on the continuum, for a recent discussion, see Ref.~\cite{roepke_pre_19}. 
%
A more systematic approach to the bound state and continuum problems has to use the Bethe-Salpeter equation (BSE), e.g.~\cite{green-book, seidel_pre_95,roepke_pre_19}. However, until now the BSE could only be solved within rather rigid approximations. The recently developed G1--G2 scheme \cite{schluenzen_prl_20,joost_prb_20,joost_prb_22} should allow for more accurate results within the GW approximation and the dynamically screened ladder approximation in the near future.

\section{Summary and Discussion} \label{s:dis}
The properties of partially ionized warm dense matter at finite temperature and high densities remain an important question that is not fully solved yet. When the system is heated and/or compressed, bound states vanish one by one. At low temperatures, this process is expected to be an insulator to metal transition that proceeds in the liquid phase and is associated with a change of the electronic structure, for a recent overview, see Ref.~\cite{bonitz_pop_24}. On the other hand, in the gas phase, at temperatures low compared to the atomic binding energy, $k_BT\ll E_1^{\rm Z}$, pressure ionization of single atoms dominates. A mix of Coulomb interaction effects (screening) and quantum and exchange effects destabilize bound states and delocalize electrons such that bound states gradually vanish one by one. Eventually, at the Mott density, $n_M$, also the ground state disappears. This density is associated to a critical value $r^{\rm Mott}_s(T^Z_R,Z)$, of the electronic coupling parameter  -- the ratio of the mean electron-electron distance to the extension of the atomic ground state wave function, and the temperature is given in units of the binding energy, $k_B T_R^Z:=E_1^Z$. This critical value is expected to increase with temperature and, when $k_BT$ approaches $E_1^Z$, bound states vanish at any density. Thus, the lowest value of $r^{\rm Mott}_s$ should be observed in the low temperature limit. 

Currently the values of $r^{\rm Mott}_s(T_R^Z,Z)$ are not known precisely. 
A value $r^{\rm Mott}_s(T=0)=1.2$ was predicted for hydrogen, based on fermionic PIMC simulations \cite{bonitz_prl_5} which, however, are severely hampered by the sign problem. Rogers \textit{et al.} \cite{rogers_pra_70} reported, for hydrogen, $r^{\rm Mott}_s=1.19$, from solutions of the Schrödinger equation with a Debye potential which reflects the influence of screening. On the other hand, atoms will break up even in a medium of non-interacting quantum particles, due to wave function overlap and  exchange effects. This will occur in a Fermi gas, if the mean kinetic energy of the electrons approaches the binding energy, $\frac{3}{5}E_F = E_R^Z$, which leads to $r_s=1.49$. Finally, experiments by Knudson \textit{et al.} on the compression of liquid deuterium \cite{knudson_science_15} detected the metal-insulator transition at $r_s=1.37$, which is quite close to the estimates above. However, this comparison has to be taken with care because the physical mechanisms in the condensed and plasma phases are different.

In this paper, we presented novel new first-principles results for the Mott density of hydrogen, that are based on fermionic PIMC simulations for hydrogen \cite{filinov_pre_23}. The results for $r_s^{\rm Mott}$ were obtained by extrapolating the effective ground state binding energy to higher density, cf. Fig.~\ref{fig:ipd_ieff}, where the fermion sign problem currently prevents direct simulations. We obtained reliable results for two temperatures: $r_s^{\rm Mott}(0.2,Z=1)\approx 1.35\dots 1.8$ and $r_s^{\rm Mott}(0.4,Z=1)\approx 1.5\dots 2$, whereas for $k_BT_R=0.1$, no prediction is possible and the behavior at $k_BT_R=0.8$ requires further analysis.

How about the $Z$-dependence of the critical value, $r_s^{\rm Mott}$, for hydrogen-like ions? We have argued that this value (for the electrons) should be $Z$-independent because it reflects wave function overlap between neighboring electrons. While this is still an open question (since the relative importance of Coulomb correlations and quantum and exchange effects varies with $Z$), this would mean that the electron density at the Mott point of different ions scales as $Z^3$, whereas the mean ion density scales, approximately, as $Z^2$. Recent beryllium compression experiments by Doeppner \textit{et al.} \cite{doeppner_nature_23} reported K-shell ``modification'' that starts in a finite density range between two data points: the first has a free electron density $n^*_e\approx 1.9\cdot 10^{24}$ cm$^{-3}$, at a temperature  of $119$eV, corresponding to $T_R^Z=0.55$, and an average charge of $\bar Z=3.05$. This corresponds to a coupling parameter of free electrons of $r^{*}_s \sim1.8$. The second point had $n^*_e\approx  7.6\cdot 10^{24}$ cm$^{-3}$, at $T=160$eV ($T_R^Z=0.74$), and $\bar Z=3.4$, corresponding to $r^{*}_s \sim 1.06$. These values are unexpectedly low, taking into account the high temperature and that we would expect the Mott density to be associated to full K-shell ionization, i.e., $\bar Z=4$. Additional high-pressure experiments, combined with first-principles simulations, are needed to verify the consistency of the reported parameters (in particular of $\bar Z$) in order to further reduce the remaining uncertainties of the Mott density across materials with different $Z$, which is important for future modeling of WDM.
In this paper we presented  a first-principles approach to the ionization potential depression in warm dense matter. The concept was first formulated in Ref.~\cite{bonitz_pop_24} and was here explored in much more detail. The main idea is to use first-principles input for the computation of the continuum lowering and IPD. In our approach, this input are the fractions of free and bound particles, $\alpha$ and $x_A$. Even though these quantities are not strictly defined, in a physical approach, such as PIMC, it is no problem to formulate clear and controlled criteria for $\alpha$ and $x_A$. As we have shown, it is straightforward to vary these criteria within a reasonable range and test the input of this variation on the IPD and establish an uncertainty interval for the result. In the present case, the variation  of the IPD, due to the criterion, did not exceed $5\%$, cf. Sec.~\ref{ss:sensitivity}.

We have presented numerical results for partially ionized hydrogen, based on fermionic PIMC results of Ref.~\cite{filinov_pre_23} for two versions: in the first, FPIMC1, no renormalization of the bound state levels was performed, whereas in FPIMC2, a finite renormalization, $\Delta_{nl}$, was taken into account. 
Since FPIMC does not provide data for $\Delta_{nl}$, we used approximate results that were obtained from solving the bound state Schrödinger equation with a Debye-type statically screened Coulomb potential of Ref.~\cite{rogers_pra_70}. 
Due to a lowering of the bound state energies in a plasma, FPIMC2 yielded significantly stronger IPD (cf. Fig.~\ref{fig:dI_allt})  and lower values for the effective ground state binding energy (cf. Fig.~\ref{fig:ipd_ieff}) than FPIMC1.

The FPIMC2-results are expected to be reliable for low electron densities, which is the situation we are restricted to anyway with our QMC input, due to the fermion sign problem. 
For higher densities and, in particular, for larger $Z$ ions, improved solutions of the bound state problem,  in a dense plasma environment, e.g., based on the BSE \cite{roepke_pre_19}, are needed as input for FPIMC2. Alternatively, the first-principles approach based on the FPIMC1-strategy can be used, and its results for the IPD will be accurate for moderate temperatures, $k_BT \ll E_R^Z$. At higher temperatures, excited state contributions will affect the results for the IPD. However, this effect is reduced by a partial compensation of correlation effects (leading to a lowering of the bound state energy eigenvalues, $E_{nl}$) and Fermi statistics (Pauli blocking, leading to an upshift of $E_{nl}$ \cite{zimmermann_pssb_78, roepke_pre_19}). One way to avoid the fermion sign problem at high densities around the Mott density would be configuration PIMC simulations \cite{schoof_cpp11,schoof_prl15}. 

Another suitable first principles approach are DFT simulations combined with reliable estimates for the degree of ionization, as was discussed in Ref.~\cite{bonitz_pop_24}. We underline that our first-principles approach to IPD can also be used with input produced by other methods. As an example, we showed in Fig.~\ref{fig:rpimc-comparison} results derived from restricted PIMC (RPIMC) simulations for hydrogen of Militzer \textit{et al}.~\cite{Militzer_PRE_2001} which we compare to the results of the present paper. Due to the significantly larger values for the degree of ionization, observed by RPIMC, for all temperatures, the binding energy $I_1^{\rm eff}$ is significantly below the FPIMC-results. 

We also pointed out an additional effect that is important for the ionization equilibrium in warm dense matter that was called over the Fermi barrier ionization. It is analogous to above threshold ionization in a strong laser field where the energy needed to liberate an electron from a nucleus is larger than the binding energy. Similarly, for ionization in  a dense plasma, electrons need to overcome not only the effective binding energy but also the Fermi barrier provided by the Pauli blocking of the ``spectator'' electrons in the plasma. While this effect is included in most treatments of the Saha equation for nonideal plasmas via the ideal chemical potential of fermions, it is important for a correct identification of the IPD and for a proper interpretation of experimental results. The simulations revealed that, for hydrogen, the Fermi barrier effect is small, due to the low Mott density. In contrast, for atoms with higher $Z$, that are of interest in current experiments, such as beryllium and carbon, the effect is expected to be relevant for comparisons with experiments.

Indeed, in recent years there has been impressive progress in experiments on the ionization of dense plasmas where indirect measurements of IPD and continuum lowering were reported. The ideal experiment to directly measure the effective ground state ionization potential $I_{1s}^{\rm eff,F}$ would be photoabsorption with a continuously variable x-ray photon energy. The threshold photon energy required for ionization would directly yield $I_{1s}^{\rm eff,F}$. However, a direct detection of photoelectrons in a plasma is challenging. Also, a broadband x-ray source is usually not available. A possible alternative are ultrashort x-ray FEL sub-spikes that deliver a broad energy spectrum, as was demonstrated in Ref.~\cite{mercadier_np_24}. However, in most recent experiments with free electron lasers only a finite set of x-ray photon energies were available, e.g., \cite{vinko_nat_12,vinko_nat-com_14}, what provides only limited energy resolution for the IPD. Instead of detecting photoelectrons, a promising diagnostic tool for the K edge are x-ray photons created by  luminescence, e.g., by electron transitions between the L- and K-shell. This has been successfully demonstrated by Vinko and co-workers \cite{vinko_nat_12,vinko_nat-com_14, ciricosta_prl_12, ciricosta_ncom_16} and works well for low and moderate compression. 
However, when the density increases and the distance between the ions approaches the extension of the bound orbitals (e.g. L- and even K-shell), bound electrons are becoming increasingly delocalized, as observed in Ref.~\cite{doeppner_nature_23}. Eventually, the L-shell will merge into the continuum and the luminescence line will disappear. A more promising and versatile tool that is applicable at arbitrary compression is x-ray Thomson scattering. However, this does not yield direct information on IPD or continuum lowering but requires model input for interpretation of the complex measurement signal. Therefore, first-principles results for the XRTS spectra, such as from QMC or DFT simulations which allow one to deduce the IPD, are particularly important.

One way how to extract the IPD from QMC data was discussed in the present paper.
Here we concentrated on simple situations where only one (hydrogen) or two-ion species (hydrogen-like ions) are present. The motivation for this was to simplify the ionization balance as much as possible so that the first-principles QMC results can be directly mapped on a nonideal Saha equation. By this ``QMC-downfolding'' procedure it is possible to determine, at least, in principle, the quantity that is left open in the nonideal Saha equation -- the IPD. This way, first-principles QMC-input can be used to derive input quantities for chemical models and to benchmark simpler approximations for the continuum shift and ionization potential depression. These benchmarks are expected to be useful also for more complex situations in warm dense matter where the plasma contains a mixture of large number of ionic species which require complex chemical models \cite{yuan_pre_25} or DFT simulations.

 \section*{Acknowledgments}
This work has been supported by the Deutsche Forschungsgemeinschaft via project BO1366-13/2.

\section*{References}

\bibliographystyle{unsrt}

\begin{thebibliography}{10}

\bibitem{ecker-kroell_63}
G.~Ecker and W.~Kröll.
\newblock {Lowering of the Ionization Energy for a Plasma in Thermodynamic Equilibrium}.
\newblock {\em The Physics of Fluids}, 6(1):62--69, 01 1963.

\bibitem{rompe-steenbeck_39}
R.~Rompe and M.~Steenbeck.
\newblock {\em {The plasma state of gases (in German)}}, volume~18 of {\em {Ergebnisse der Exakten Naturwissenschaften}}.
\newblock 1939.

\bibitem{inglis-teller_39}
D.~R. {Inglis} and E.~{Teller}.
\newblock {Ionic Depression of Series Limits in One-Electron Spectra.}
\newblock {\em Astrophys. Journal}, 90:439, October 1939.

\bibitem{stewart-pyatt_66}
John~C. {Stewart} and Jr. {Pyatt}, Kedar~D.
\newblock {Lowering of Ionization Potentials in Plasmas}.
\newblock {\em \apj}, 144:1203, June 1966.

\bibitem{crowley_pra_90}
B.~J.~B. Crowley.
\newblock Average-atom quantum-statistical cell model for hot plasma in local thermodynamic equilibrium over a wide range of densities.
\newblock {\em Phys. Rev. A}, 41:2179--2191, Feb 1990.

\bibitem{preston_hedp_13}
Thomas~R. Preston, Sam~M. Vinko, Orlando Ciricosta, Hyun-Kyung Chung, Richard~W. Lee, and Justin~S. Wark.
\newblock The effects of ionization potential depression on the spectra emitted by hot dense aluminium plasmas.
\newblock {\em High Energy Density Physics}, 9(2):258--263, 2013.

\bibitem{ciricosta_prl_12}
O.~Ciricosta, S.~M. Vinko, H.-K. Chung, B.-I. Cho, C.~R.~D. Brown, T.~Burian, J.~Chalupsk\'y, K.~Engelhorn, R.~W. Falcone, C.~Graves, V.~H\'ajkov\'a, A.~Higginbotham, L.~Juha, J.~Krzywinski, H.~J. Lee, M.~Messerschmidt, C.~D. Murphy, Y.~Ping, D.~S. Rackstraw, A.~Scherz, W.~Schlotter, S.~Toleikis, J.~J. Turner, L.~Vysin, T.~Wang, B.~Wu, U.~Zastrau, D.~Zhu, R.~W. Lee, P.~Heimann, B.~Nagler, and J.~S. Wark.
\newblock Direct measurements of the ionization potential depression in a dense plasma.
\newblock {\em Phys. Rev. Lett.}, 109:065002, Aug 2012.

\bibitem{yuan_pre_25}
Jiaolong Zeng, Aihua Deng, Cheng Gao, Yong Hou, and Jianmin Yuan.
\newblock Extended chemical picture of ionization balance to extremely dense plasmas.
\newblock {\em Phys. Rev. E}, 111:015211, Jan 2025.

\bibitem{zimmermann_pssb_78}
R.~Zimmermann, K.~Kilimann, W.~D. Kraeft, D.~Kremp, and G.~Röpke.
\newblock Dynamical screening and self-energy of excitons in the electron–hole plasma.
\newblock {\em physica status solidi (b)}, 90(1):175--187, 1978.

\bibitem{haug_pqe_84}
H.~Haug and S.~Schmitt-Rink.
\newblock Electron theory of the optical properties of laser-excited semiconductors.
\newblock {\em Progress in Quantum Electronics}, 9(1):3--100, 1984.

\bibitem{red-book}
Werner Ebeling, Wolf Kraeft, and Dietrich Kremp.
\newblock {\em Theory of Bound States and Ionization Equilibrium in Plasmas and Solids}.
\newblock Akademie-Verlag, Berlin, 01 1976.

\bibitem{roepke_pre_19}
Gerd R\"opke, David Blaschke, Tilo D\"oppner, Chengliang Lin, Wolf-Dietrich Kraeft, Ronald Redmer, and Heidi Reinholz.
\newblock Ionization potential depression and pauli blocking in degenerate plasmas at extreme densities.
\newblock {\em Phys. Rev. E}, 99:033201, Mar 2019.

\bibitem{vinko_nat_12}
S.~M. Vinko, O.~Ciricosta, B.~I. Cho, K.~Engelhorn, H.-K. Chung, C.~R.~D. Brown, T.~Burian, J.~Chalupsk{\'y}, R.~W. Falcone, C.~Graves, V.~H{\'a}jkov{\'a}, A.~Higginbotham, L.~Juha, J.~Krzywinski, H.~J. Lee, M.~Messerschmidt, C.~D. Murphy, Y.~Ping, A.~Scherz, W.~Schlotter, S.~Toleikis, J.~J. Turner, L.~Vysin, T.~Wang, B.~Wu, U.~Zastrau, D.~Zhu, R.~W. Lee, P.~A. Heimann, B.~Nagler, and J.~S. Wark.
\newblock Creation and diagnosis of a solid-density plasma with an x-ray free-electron laser.
\newblock {\em Nature}, 482(7383):59--62, Feb 2012.

\bibitem{kasim_np_18}
M.~F. Kasim, J.~S. Wark, and S.~M. Vinko.
\newblock Validating continuum lowering models via multi-wavelength measurements of integrated x-ray emission.
\newblock {\em Scientific Reports}, 8(1):6276, Apr 2018.

\bibitem{ciricosta_ncom_16}
O.~Ciricosta, S.~M. Vinko, B.~Barbrel, D.~S. Rackstraw, T.~R. Preston, T.~Burian, J.~Chalupsk{\'y}, B.~I. Cho, H.-K. Chung, G.~L. Dakovski, K.~Engelhorn, V.~H{\'a}jkov{\'a}, P.~Heimann, M.~Holmes, L.~Juha, J.~Krzywinski, R.~W. Lee, S.~Toleikis, J.~J. Turner, U.~Zastrau, and J.~S. Wark.
\newblock Measurements of continuum lowering in solid-density plasmas created from elements and compounds.
\newblock {\em Nature Communications}, 7(1):11713, May 2016.

\bibitem{doeppner_nature_23}
T.~D{\"o}ppner, M.~Bethkenhagen, D.~Kraus, P.~Neumayer, D.~A. Chapman, B.~Bachmann, R.~A. Baggott, M.~P. B{\"o}hme, L.~Divol, R.~W. Falcone, L.~B. Fletcher, O.~L. Landen, M.~J. MacDonald, A.~M. Saunders, M.~Sch{\"o}rner, P.~A. Sterne, J.~Vorberger, B.~B.~L. Witte, A.~Yi, R.~Redmer, S.~H. Glenzer, and D.~O. Gericke.
\newblock Observing the onset of pressure-driven k-shell delocalization.
\newblock {\em Nature}, 618(7964):270--275, Jun 2023.

\bibitem{mercadier_np_24}
Laurent Mercadier, Andrei Benediktovitch, {\v{S}}pela Kru{\v{s}}i{\v{c}}, Joshua~J. Kas, Justine Schlappa, Marcus Ag{\aa}ker, Robert Carley, Giuseppe Fazio, Natalia Gerasimova, Young~Yong Kim, Lo{\"i}c Le~Guyader, Giuseppe Mercurio, Sergii Parchenko, John~J. Rehr, Jan-Erik Rubensson, Svitozar Serkez, Michal Stransky, Martin Teichmann, Zhong Yin, Matja{\v{z}} {\v{Z}}itnik, Andreas Scherz, Beata Ziaja, and Nina Rohringer.
\newblock Transient absorption of warm dense matter created by an x-ray free-electron laser.
\newblock {\em Nature Physics}, 20(10):1564--1569, Oct 2024.

\bibitem{militzer_prr_22}
G.~Massacrier, M.~B\"ohme, J.~Vorberger, F.~Soubiran, and B.~Militzer.
\newblock Reconciling ionization energies and band gaps of warm dense matter derived with ab initio simulations and average atom models.
\newblock {\em Phys. Rev. Res.}, 3:023026, Apr 2021.

\bibitem{santra_pre_22}
Rui Jin, Zoltan Jurek, Robin Santra, and Sang-Kil Son.
\newblock Plasma environmental effects in the atomic structure for simulating x-ray free-electron-laser-heated solid-density matter.
\newblock {\em Phys. Rev. E}, 106:015206, Jul 2022.

\bibitem{huang_pre_24}
Yihua Huang, Zhenhao Liang, Jiaolong Zeng, and Jianmin Yuan.
\newblock Nonideal effects on ionization potential depression and ionization balance in dense al and au plasmas.
\newblock {\em Phys. Rev. E}, 109:045210, Apr 2024.

\bibitem{davletov_njp_23}
A~E Davletov, Yu~V Arkhipov, Ye~S Mukhametkarimov, L~T Yerimbetova, and I~M Tkachenko.
\newblock Generalized chemical model for plasmas with application to the ionization potential depression.
\newblock {\em New Journal of Physics}, 25(6):063019, jun 2023.

\bibitem{hu_prl_17}
S.~X. Hu.
\newblock Continuum lowering and fermi-surface rising in strongly coupled and degenerate plasmas.
\newblock {\em Phys. Rev. Lett.}, 119:065001, Aug 2017.

\bibitem{vinko_nat-com_14}
S.~M. Vinko, O.~Ciricosta, and J.~S. Wark.
\newblock Density functional theory calculations of continuum lowering in strongly coupled plasmas.
\newblock {\em Nature Communications}, 5(1):3533, Mar 2014.

\bibitem{filinov_pre_23}
Alexey Filinov and Michael Bonitz.
\newblock The equation of state of partially ionized hydrogen and deuterium plasma revisited.
\newblock {\em Phys. Rev. E}, {\bf 108}:055212, 2023.

\bibitem{bonitz_pop_24}
M.~Bonitz, J.~Vorberger, M.~Bethkenhagen, M.~Böhme, D.M. Ceperley, A.~Filinov, T.~Gawne, F.~Graziani, G.~Gregori, P.~Hamann, S.B. Hansen, M.~Holzmann, S.X. Hu, H.~Kählert, V.~Karasiev, U.~Kleinschmidt, L.~Kordts, C.~Makait, B.~Militzer, Z.A. Moldabekov, C.~Pierleoni, M.~Preising, K.~Ramakrishna, R.~Redmer, S.~Schwalbe, P.~Svensson, and T.~Dornheim.
\newblock {Towards First-principles based simulations of dense hydrogen}.
\newblock {\em Phys. Plasmas}, 31(11):110501, 2024.

\bibitem{saha_20}
Megh~Nad Saha.
\newblock Liii. ionization in the solar chromosphere.
\newblock {\em The London, Edinburgh, and Dublin Philosophical Magazine and Journal of Science}, 40(238):472--488, 1920.

\bibitem{schlanges-etal.95cpp}
M.~Schlanges, M.~Bonitz, and A.~Tschttschjan.
\newblock Plasma phase transition in fluid hydrogen--helium mixtures.
\newblock {\em Contrib. Plasma Phys.}, {\bf 35}:109, 1995.

\bibitem{blue-book}
D.~Kremp, M.~Schlanges, and W.-D. Kraeft.
\newblock {\em Quantum Statistics of Nonideal Plasmas}.
\newblock Springer, Heidelberg, 2005.

\bibitem{Note1}
Note that in order to include this contribution into all energy levels, we first have to go back to the original partition function and make the transition to the Planck-Larkin form at the final step.

\bibitem{bonitz_qkt}
M.~Bonitz.
\newblock {\em {Quantum Kinetic Theory}}.
\newblock Springer, Cham, 2 edition, 2016.

\bibitem{kremp_99_pre}
D.~Kremp, Th. Bornath, M.~Bonitz, and M.~Schlanges.
\newblock Quantum kinetic theory of plasmas in strong laser fields.
\newblock {\em Phys. Rev. E}, {\bf 60}:4725--4732, Oct 1999.

\bibitem{bonitz_99_cpp}
M.~Bonitz, Th. Bornath, D.~Kremp, M.~Schlanges, and W.~D. Kraeft.
\newblock {Quantum Kinetic Theory for Laser Plasmas. Dynamical Screening in Strong Fields}.
\newblock {\em Contrib. Plasma Phys.}, {\bf 39}(4):329--347, 1999.

\bibitem{agostini_prl_79}
P.~Agostini, F.~Fabre, G.~Mainfray, G.~Petite, and N.~K. Rahman.
\newblock Free-free transitions following six-photon ionization of xenon atoms.
\newblock {\em Phys. Rev. Lett.}, 42:1127--1130, Apr 1979.

\bibitem{becker_aamop_02}
W.~Becker, F.~Grasbon, R.~Kopold, D.B. Milošević, G.G. Paulus, and H.~Walther.
\newblock Above-threshold ionization: From classical features to quantum effects.
\newblock volume~48 of {\em Advances In Atomic, Molecular, and Optical Physics}, pages 35--98. Academic Press, 2002.

\bibitem{agostini_jpb_24}
Pierre Agostini.
\newblock What future for attophysics?
\newblock {\em Journal of Physics B: Atomic, Molecular and Optical Physics}, 57(16):162501, jul 2024.

\bibitem{Note2}
Due to the finite number of bound states in the presence of the plasma (except for the zero density limit) the canonical partition function is convergent and the Planck-Larkin procedure is not required. However, keeping the latter has the advantage that the partition function is a continuous function of density.

\bibitem{ebeling-richert_85}
W.~Ebeling and W.~Richert.
\newblock Plasma phase transition in hydrogen.
\newblock {\em Physics Letters A}, 108(2):80--82, 1985.

\bibitem{2017Ebeling}
W.~Ebeling, V.~E. Fortov, and V.~S. Filinov.
\newblock {\em {Quantum statistics of dense gases and nonideal plasmas}}.
\newblock Springer series in plasma science and technology. Springer, Cham, 2017.

\bibitem{rogers_pra_70}
F.~J. Rogers, H.~C. Graboske, and D.~J. Harwood.
\newblock Bound eigenstates of the static screened coulomb potential.
\newblock {\em Phys. Rev. A}, 1:1577--1586, Jun 1970.

\bibitem{seidel_pre_95}
J.~Seidel, S.~Arndt, and W.~D. Kraeft.
\newblock Energy spectrum of hydrogen atoms in dense plasmas.
\newblock {\em Phys. Rev. E}, 52:5387--5400, Nov 1995.

\bibitem{green-book}
W.-D. Kraeft, D.~Kremp, W.~Ebeling, and G.~R\"opke.
\newblock {\em Quantum Statistics of Charged Particle Systems}.
\newblock Akademie-Verlag, Berlin, 1986.

\bibitem{onate_jtap_16}
C.~A. Onate and J.~O. Ojonubah.
\newblock {Eigensolutions of the Schr{\"o}dinger equation with a class of Yukawa potentials via supersymmetric approach}.
\newblock {\em Journal of Theoretical and Applied Physics}, 10(1):21--26, Mar 2016.

\bibitem{bonitz_prl_5}
M.~Bonitz, V.~S. Filinov, V.~E. Fortov, P.~R. Levashov, and H.~Fehske.
\newblock {Crystallization in Two-Component Coulomb Systems}.
\newblock {\em Phys. Rev. Lett.}, {\bf 95}:235006, Dec 2005.

\bibitem{Militzer_PRE_2001}
B.~Militzer and D.~M. Ceperley.
\newblock Path integral monte carlo simulation of the low-density hydrogen plasma.
\newblock {\em Phys. Rev. E}, 63:066404, May 2001.

\bibitem{schluenzen_prl_20}
Niclas Schlünzen, Jan-Philip Joost, and Michael Bonitz.
\newblock {Achieving the Scaling Limit for Nonequilibrium Green Functions Simulations}.
\newblock {\em Phys. Rev. Lett.}, {\bf 124}(7):076601, 2020.

\bibitem{joost_prb_20}
Jan-Philip Joost, Niclas Schl\"unzen, and Michael Bonitz.
\newblock {G1-G2 scheme: Dramatic acceleration of nonequilibrium Green functions simulations within the Hartree-Fock generalized Kadanoff-Baym ansatz}.
\newblock {\em Phys. Rev. B}, {\bf 101}:245101, Jun 2020.

\bibitem{joost_prb_22}
J.-P. Joost, N.~Schl\"unzen, H.~Ohldag, M.~Bonitz, F.~Lackner, and I.~Brezinova.
\newblock The dynamically screened ladder approximation: Simultaneous treatment of strong electronic correlations and dynamical screening out of equilibrium.
\newblock {\em Physical Review B}, {\bf 105}:165155, 2022.

\bibitem{knudson_science_15}
M.~D. Knudson, M.~P. Desjarlais, A.~Becker, R.~W. Lemke, K.~R. Cochrane, M.~E. Savage, D.~E. Bliss, T.~R. Mattsson, and R.~Redmer.
\newblock Direct observation of an abrupt insulator-to-metal transition in dense liquid deuterium.
\newblock {\em Science}, 348(6242):1455--1460, 2015.

\bibitem{schoof_cpp11}
T.~Schoof, M.~Bonitz, A.~Filinov, D.~Hochstuhl, and J.W. Dufty.
\newblock Configuration path integral {M}onte {C}arlo.
\newblock {\em Contrib. Plasma Phys.}, {\bf 51}:687--697, 2011.

\bibitem{schoof_prl15}
T.~Schoof, S.~Groth, J.~Vorberger, and M.~Bonitz.
\newblock \textit{Ab Initio} thermodynamic results for the degenerate electron gas at finite temperature.
\newblock {\em Phys. Rev. Lett.}, {\bf 115}:130402, 2015.

\end{thebibliography}

%


\end{document}